\documentclass[%
preprint,
superscriptaddress,
showpacs,preprintnumbers,
bibnotes,
 amsmath,amssymb,
 aps,
pre,
]{revtex4-1}

\usepackage[T1]{fontenc}
\usepackage{mathptmx}
\usepackage{natbib}

\usepackage{graphicx}
\usepackage{dcolumn}
\usepackage{subfigure}
\usepackage{epsfig}
\usepackage{epstopdf}
\usepackage{bm}
\usepackage[mathlines]{lineno}

\begin{document}

\newcommand{\reffig}[1]{Fig.\ref{#1}}
\newcommand{\equref}[1]{Eq. (\ref{#1})}
\newcommand{\ep}{$E_p$ }
\newcommand{\et}{$E_t$ }
\newcommand{\eb}{$E_b$ }
\newcommand{\al}{\alpha }
\newcommand{\ka}{\kappa }
\newcommand{\om}{\omega }
\newcommand{\ta}{\tau }
\newcommand{\be}{\beta }
\newcommand{\pa}{\partial }

\preprint{APS/123-QED}

\title{Formation of Non-uniform Double Helices for Elastic Rods under Torsion}

\author{Hongyuan Li}
\author{Shumin Zhao}
\email{zhaosm@mail.xjtu.edu.cn}
\affiliation{Department of Applied Physics, School of Science, Xi'an Jiaotong University, Shaanxi 710049, People's Republic of China}

\author{Minggang Xia}
\affiliation{Department of Applied Physics, School of Science, Xi'an Jiaotong University, Shaanxi 710049, People's Republic of China}
\affiliation{Department of Optical Information Science and Technology, School of Science, Xi'an Jiaotong University, 710049, People's Republic of China}
\affiliation{Laboratory of Nanostructure and Physics Properties, School of Science, Xi'an Jiaotong University, 710049, People's Republic of China}
\author{Siyu He}
\affiliation{Department of Applied Physics, School of Science, Xi'an Jiaotong University, Shaanxi 710049, People's Republic of China}
\author{Qifan Yang}
\affiliation{Department of Biomedical Engineering, School of Life Science and Technology, Xi'an Jiaotong University, Shaanxi 710049, People's Republic of China}
\author{Yuming Yan}
\affiliation{Department of Electrical Engineering and Automation, School of Electrical Engineering, Xi'an Jiaotong University, Shaanxi 710049, People's Republic of China}
\author{Hanqiao Zhao}
\affiliation{Department of Biomedical Engineering, School of Life Science and Technology, Xi'an Jiaotong University, Shaanxi 710049, People's Republic of China}

\date{\today}

\begin{abstract}
The spontaneous formation of double helices for filaments under torsion is common and significant. For example, the research on the supercoiling of DNA is helpful for understanding the replication and transcription of DNA. Similar double helices can appear in polymer chains, carbon nanotube yarns, cables, telephone wires and so forth. 
We noticed that non-uniform double helices can be produced due to the surface friction induced by the self-contact. 
Therefore an ideal model was presented to investigate the formation of double helices for elastic rods under torque. 
A general equilibrium condition which is valid for both the smooth surface and the rough surface situations is derived by using the variational method. 
Based on this, by adding further constraints, the smooth and rough surface situations are investigated in detail respectively. 
Additionally, the model showed that the specific process of how to twist and slack the rod can determine the surface friction and hence influence the configuration of the double helix formed by rods with rough surfaces. Based on this principle, a method of manufacturing double helices with designed configurations was proposed and demonstrated.  Finally, experiments were performed to verify the model and the results agreed well with the theory.

\end{abstract}


\maketitle

\section{\label{sec:level1}Introduction}
It is commonly observed that a twisted yarn tends to writhe and form a double-helical structures. The formation of double helices for twisted filaments is of great significance and it has attracted attention from various fields. It is known that the supercoiling of DNA chains plays an important role in the replication and transcription of DNA \cite{shape_of_supercoiled}. The formation of supercoiling for DNA chains under torsion have been investigated broadly \cite{twisting_DNA1,twisting_DNA2,Strick1998,Strick2003,braiding_DNA1}. Besides, some biosensors utilizing supercoiled DNA chains have been reported \cite{biosensor,biosensor2}.  
Y. Shang \emph{et al.} obtained the double helices of carbon nano-tube (CNT) yarns by simply twisting and slacking it, which have unique electrical and mechanical properties \cite{CNT1,CNT2,CNT3}. The superconductor made from the double helix of CNT yarns have been investigated broadly \cite{gao2016}. Besides, the double helix of carbon or polymer fibers can be fabricated into the artificial muscle, which, as a new concept, have recently attracted wide attention \cite{Haines2014,Yue2015}.
In engineering, cables under torsion may form loops and double helices, in which the large deformation may damage the cables \cite{loop_formation_DNA_Cable,cable_DNA_writhe}.

Two main methods have been proposed to investigate the spatial writhing of twisted filaments. The first is to simplify the filament or polymer chain as an ideal elastic rod. Based on this elastic-rod assumption, the initial buckling and localized post-buckling, prior to self-contact, of a twisted elastic rod have been widely investigated by using the Kirchhoff model \cite{experiment1,experiment2,experiment3,theory_Kirchhoff,theorypurohit}. The self-contact for the two strands of double helices were studied by some scholars as well \cite{self_contact1,self_contact2,self_contact3,self_contact4}. The other method mainly focuses on microscopic polymer chains. To describe the statistical property of polymer chains, the Kratky-Porod model, free jointed chain model and worm like chain model were put forward \cite{chainmodel}. The statistical properties of the supercoiling of DNA chains can be described well with these models \cite{DNAtorsion,elasticmodel}.

Although the self-contact is considered in some elastic-rod model, most scholars still neglected the surface friction \cite{self_contact1,theorypurohit}. Without considering the surface friction, the double helix solved by whether the Kirchhoff model or the variational method will have a uniform double helix (if the slight deviation near the ends is neglected). However, we noticed in most cases the surface friction is not negligible. On the contrary, it may have a great influence on the configuration and mechanical properties of the double helices.
A typical non-uniform double helix can be obtained by twisting a rubber rod first and then bringing its two ends together. 
In the work of Y. Shang \emph{et al.}\cite{CNT1,CNT2,CNT3}, the double helices of CNT yarns obtained by similar twist-slacking process shown non-uniformity as well.

To research the formation of double helices for rods under torque and how to control their configurations is meaningful. First, it can help us understand the mechanism of the supercoiling formation of DNA chains and other polymers. Moreover, it has potential in controlling the fabrication of the double helices of CNT yarns which can serve as high-performance electronic and mechanical devices, for example, the  superconductors and artificial muscles. Besides, in engineering, it can be used to avoid the double helix formation for cables.

In this article, based on the ideal elastic rod model, the variational method is used to derive the general equilibrium condition for the double helix. This condition is valid for both the smooth surface and the rough surface situations. By adding a inter-strand interaction term in the total potential energy, both the self-contact and non-self-contact situations are considered.
Different from previous non-friction assumptions, the influence of the surface friction is investigated here. It is found that the surface friction will greatly influence the configuration of the double helix, in which case, non-uniform double helices can be formed. Moreover, the specific process of how the rod is twisted and slacked can determine the surface friction and therefore influence the configuration of the double helix. Based on this property, a method of producing double helices with designed configurations is proposed. Experiments are preformed to verify the validity of this model and demonstrate the method of producing a double helix with a designed configuration.

\section{\label{sec:level1}Model}
\subsection{\label{sec:level2}General theory}
\begin{figure*}
\subfigure[]{
    \includegraphics[width=0.45\textwidth]{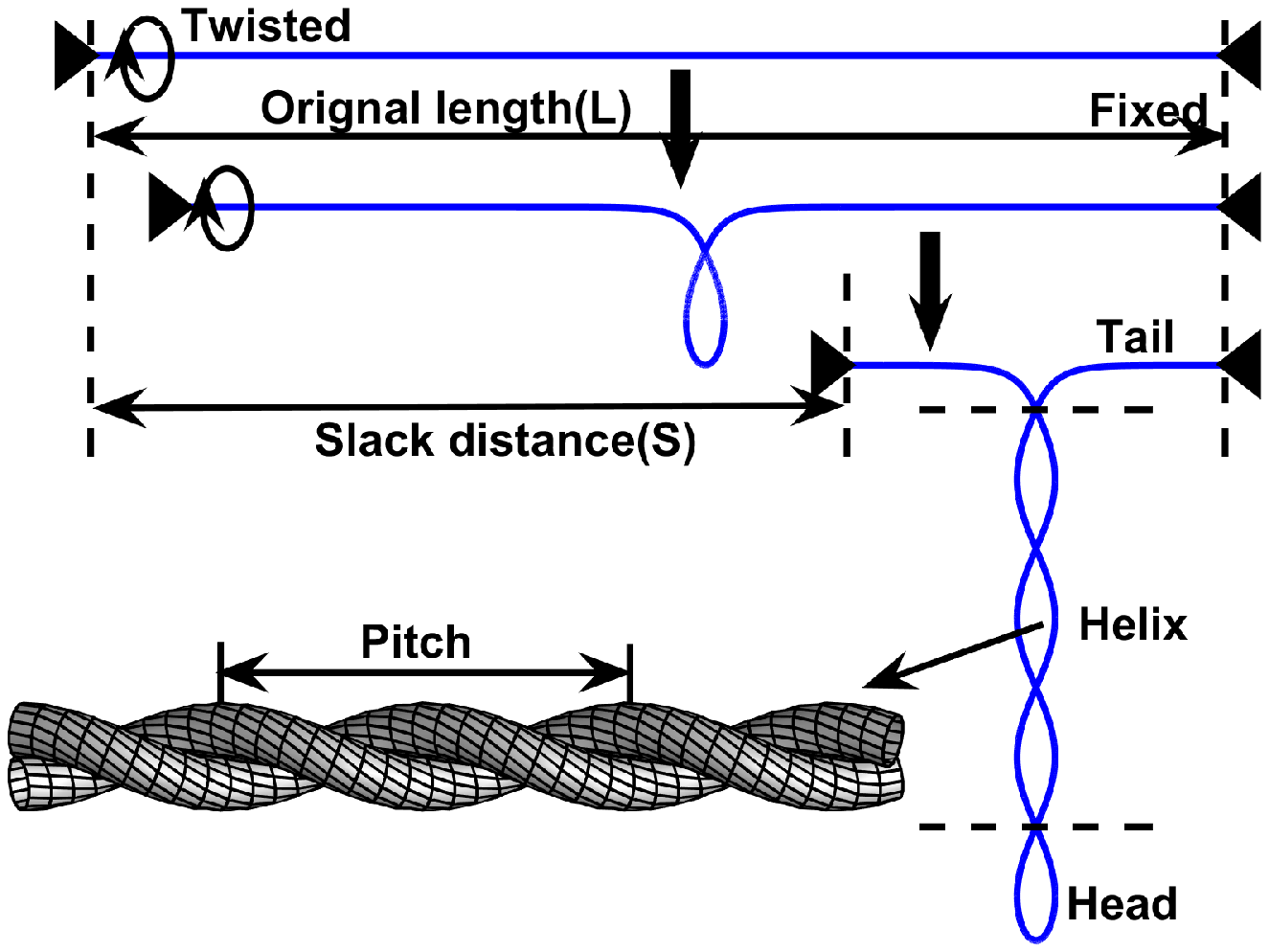}
    \label{figure1:subfig:1}
}
\subfigure[]{
    \includegraphics[width=0.45\textwidth]{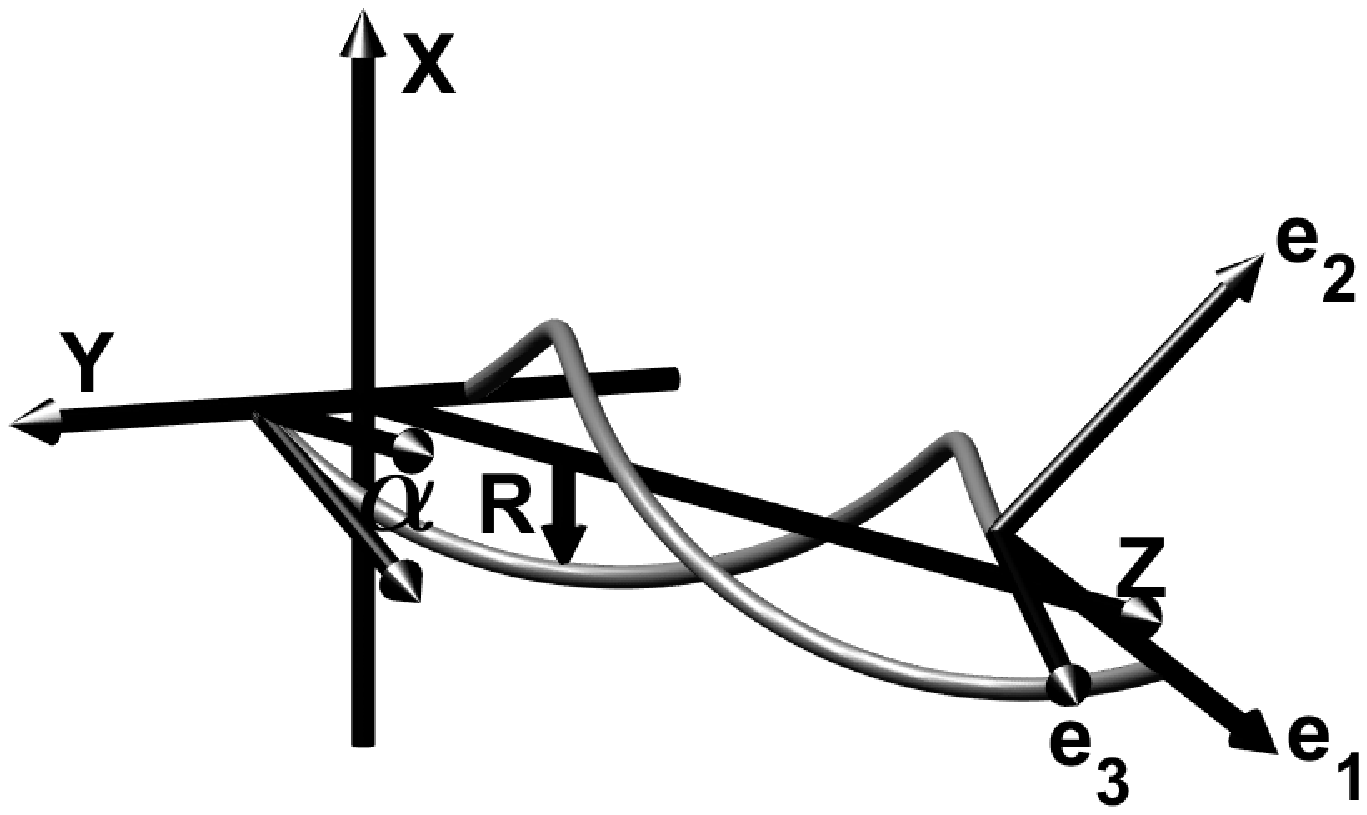}
	\label{figure1:subfig:2}
}
    \caption{(a) Schematic of the formation process of the double helix. One end is fixed while the other end can be slacked and twisted. The whole elastic rod is divided into three parts: head, helix and tail. The pitch the double helix are illustrated. (b) Central lines of the double helix. $\alpha$ is the angle between the tangent vector of the helix and the z axis. $R$ is the raidus of the double helix central lines. $\textbf{e}_3$, $\textbf{e}_1$ and $\textbf{e}_2$ are the unitary vector in the directions of $\textbf{r}'$, $\textbf{r}''$ and $\textbf{r}'\times \textbf{r}''$ respectively.}\label{figure1:subfig}
\end{figure*}

In this section, the general properties of the double helices are investigated. By using the variational approach based on the energy minimization principle, the general equilibrium condition can be obtained. \reffig{figure1:subfig:1} shows the schematic of the formation process of the double helix. Here, we call this procedure the twist-slacking method so as to distinguish it from the conventional twist-spinning method \cite{gao2016,braiding_DNA1}. To simplify the problem and obtain an ideal model, $5$ fundamental assumptions are made as follows:

\begin{itemize}
    \item 1. The rod is inextensible with a fixed length $L$. It has a circular cross section with a fixed radius $r$.
    \item 2. The elasticity of the rod is linear. The twisting and bending are uncoupled. The linear-elasticity assumption can be satisfied well in our experiments. See more details in the experiment section.
    \item 3. The rod can be roughly divided into three parts, as shown in \reffig{figure1:subfig:1}, namely the tail, helix and head. Since we mainly focus on the behavior of the helix part, the influence of the head is neglected here and the tail is assumed to be straight and uniform. This assumption can be perfectly satisfied when the helix is long enough. The radius of the double helix composed by the central line of the rod is $R$, as shown in \reffig{figure1:subfig:2}.
    \item 4. The rod is quasi-static during the whole slack process.
    \item 5. No dissipative force exists in the formation process, including the sliding friction. However static friction is allowed.
\end{itemize}

Then we describe the geometry of the double helix quantitatively. According to assumption $3$, all the slack distance $S$ can be transformed into the length of the two strands in the double helix, as illustrated in \reffig{figure1:subfig:1}. Assume the double helix extends in the positive z-axis direction so that the two strands are symmetric about the z-axis, as shown in \reffig{figure1:subfig:2}. Due to the symmetry, we only need to describe one of the two strands. The position vector can be written as
\begin{equation}\label{position vector}
{\bf{r}(\theta)} = \left( {R\cos \theta ,R\sin \theta ,\int_0^\theta \frac{P}{2\pi}d\theta } \right),
\end{equation}
where \(\theta\) defines the angle that the helix rotates by and $P$ is the pitch of the helix (shown in \reffig{figure1:subfig:1}. See the exact definition in Ref. \cite{rope_geometry}).
Here, we will use the natural coordinate $s$ with its origin located at $\textbf{r}(0)$. It represents the distance along the centre line. Then $0\leqslant s<S/2$ corresponds to the helix, and $S/2\leqslant s<L/2$ corresponds to the tail. For the helix part we have
\begin{equation}\label{ds0}
ds=d{|\bf{r}'|}=\sqrt{R^2+\left({\frac{P}{2\pi}}\right)^2}d\theta.
\end{equation}
The curvature \(\kappa\) and the torsion $\tau$ can be calculated as follows
\begin{equation}\label{curvature}
\kappa =\frac{{|\bf{r}'}\times \bf{r}''|}{{|\bf{r}'|}^3},
\tau =\frac{\left(\bf{r}'\times\bf{r}'' \right)\cdot\bf{r}'''}{|\bf{r}'\times\bf{r}''|^2},
\end{equation}
where $\textbf{r}'$, $\textbf{r}''$ and $\textbf{r}'''$ are the first, second and third order derivatives of $\textbf{r}$ with respect to $\theta$. By using \equref{position vector}, the curvature and torsion can be simplified as follows
\begin{equation}\label{Eq curvature1}
\kappa  = \frac{4{\pi}^2R}{{{(2\pi R)^2} + {P^2}}}=\frac{\sin^2\al}{R},
\end{equation}
\begin{equation}\label{Eq torsion1}
\tau=\frac{2\pi P}{{(2\pi R)^2} + {P^2}}=\frac{\cos\al\sin\al}{R},
\end{equation}
where $\al$ is defined by
\begin{equation}\label{eq alpha def}
\al=\arctan\left(\frac{2\pi R}{P}\right).
\end{equation}
It represents the tilting angle between the tangent vector of central line of the helix and the z axis, as illustrated in \reffig{figure1:subfig:2}.
Note in \equref{Eq torsion1} we use the approximation that the first and higher order derivatives of $P$ with respect to $s$ are neglected, since the variation of the pitch is very slow here. The validity of this approximation is discussed in Appendix A. 

It is noteworthy that for some microscopic systems including the DNA chains, the symmetric assumption about the two strands may be invalidated due to the thermal fluctuation. In this case, if the deviation from the double helix configuration is not too serious, we can still roughly analyze it with our model. $P$ can be redefined by the average distance between every two crossings of the double helix. $R$ can be replaced by half the average minimum distance between the two strands. Then $\ka$, $\ta$ and $\al$ can be roughly estimated by using \equref{Eq curvature1}, \equref{Eq torsion1} and \equref{eq alpha def}.

The Calugareanu invariant is introduced here to describe the topological characteristics of the continuous shape variation in the rod \cite{calugareanu_invariant,Self_link}.
\begin{equation}\label{Lk=Tw+Wr}
Lk=Tw+Wr,
\end{equation}
where $Tw$ represents the total twist of the rod around its central line and $Wr$ represents the degree of writhing of the rod, which is a property of the centre line of the rod. $Lk$ is the linking number, which is a topological invariant here. In our case, the precise expressions for \(Lk\), \(Tw\) and \(Wr\) can be expressed as \cite{Writhing_number}

\begin{equation}\label{Eq Lk=Tw+Wr}
\begin{aligned}
&Lk=N(S),\\
&Tw=2 \cdot\frac{1}{2\pi} \int_0^{L/2} \omega ds, \\
&Wr= 2 \cdot \frac{1}{2\pi} \int_0^{S/2} \tau ds,
\end{aligned}
\end{equation}
where $\omega$ is the twist per unit length, $N(S)$ is the number of turns by which the rod is rotated. Since the rod can be twisted and slacked simultaneously, $N$ can be written as a function of $S$ here.
The coefficient of 2 in $Tw$ and $Wr$ is due to there being two strands.
It is noteworthy that some scholars expressed $Wr$ approximately with the number of (signed)crossings for the double helix in just one view, for example, in the direction of the x or y axis \cite{Simplification_Tw}. According to Fuller, $Wr$ should equal to the crossing numbers averaged over all views \cite{Writhing_number}. The approximation is equivalent to replacing $\ta$ with $\sin\al/R$ in \equref{Eq torsion1}, which is only valid when $\alpha$ is very small.

\begin{figure}
  \centering
  \includegraphics[width=0.5\linewidth]{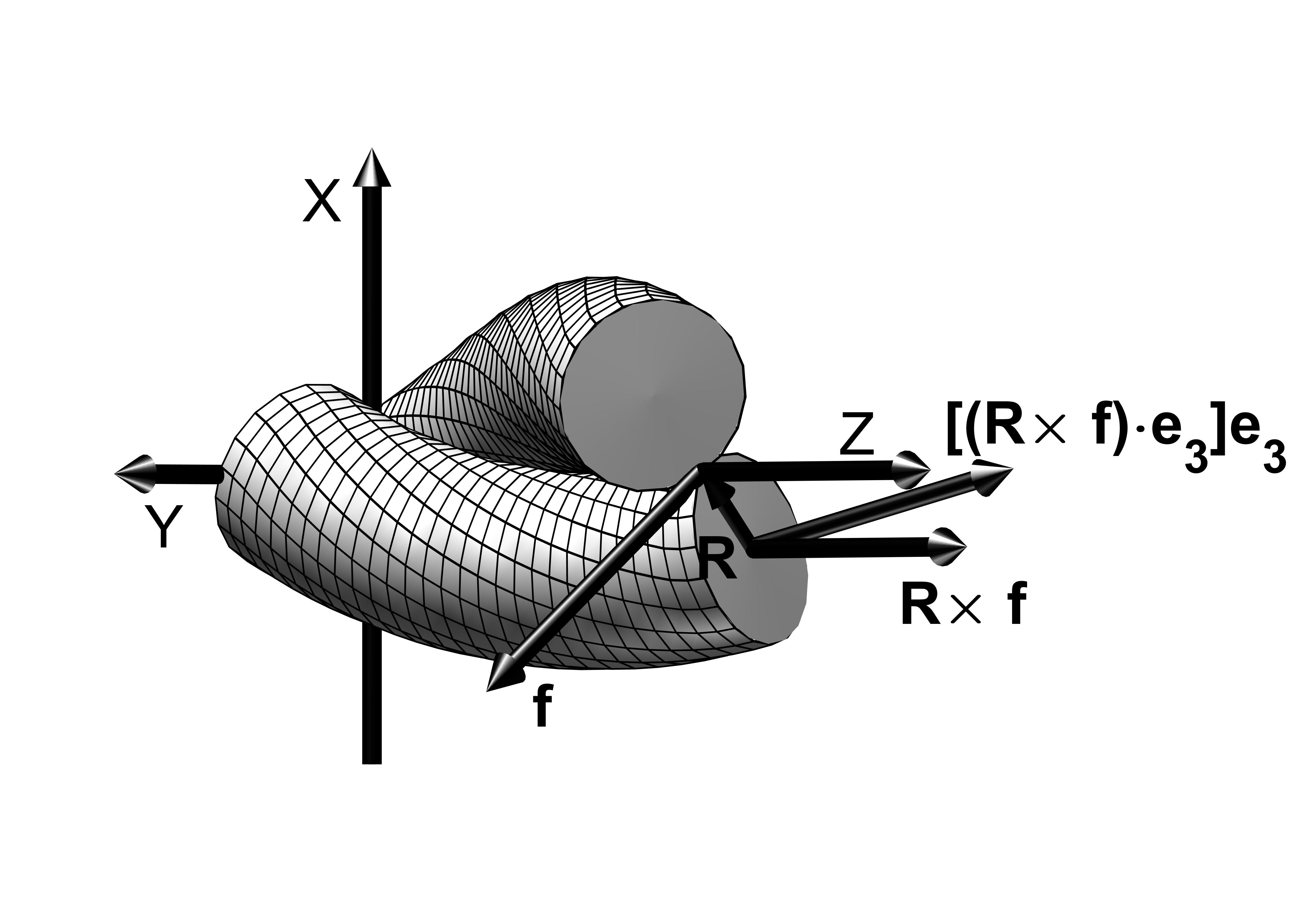}\\
  \caption{Schematic of the twist moment provided by the surface friction. The surface friction is perpendicular to the z axis due to symmetry and has a moment $\textbf{R}\times\textbf{f}$. $M_f$ is the projection of it on the direction of $\textbf{e}_3$}\label{friction1}
\end{figure}

In addition, along the central line of the rod, the equilibrium of moment requires \cite{experiment1}
\begin{equation}\label{eq twist moment der}
\frac{dM_3}{ds}=0,
\end{equation}
where $M_3$ is the third component of the moment(in the Frenet coordinate, shown in \reffig{figure1:subfig:2}), namely the twist moment. Here if we include the influence of the friction, the twist moment can be expressed as
\begin{equation}\label{eq twist moment}
M_3=C\om+M_f,
\end{equation}
where $M_f$ is the twist moment provided by the surface friction. Since the two central lines are symmetric about the z-axis, the friction must be perpendicular to the z axis, as shown in \reffig{friction1} . Then we have
\begin{equation}\label{eq dM_f}
dM_f = (\textbf{R}\times\textbf{f})\cdot \textbf{e}_3ds
=\cos\al R f ds,
\end{equation}
in which $\textbf{f}$ is the surface friction per unit length and $\textbf{e}_3$ is the unitary vector in the tangent direction of the central line. Obviously, $f$ is $0$ for the helix part, which means $d\om/ds=0$ holds for $s>S/2$. Therefore, $\omega$ must have the form

\begin{equation}\label{Eq omega}
\omega(s;S)=\left\{
            \begin{aligned}
            &\omega(s;S) \quad (0\leqslant s\leqslant S/2) \\
            &\omega(S/2;S) \quad (S/2 < s\leqslant L/2) \\
            \end{aligned}
            \right.,
\end{equation}
where the first argument is the natural coordinate and the second argument is the slack distance. For convenience, we define
\begin{equation}
\om_t=\om(S/2;S).
\end{equation}
The influence of the surface friction on the specific form of $\omega(s;S)$ will be discussed in the next two subsections.

The total potential energy can be easily written as \cite{energy}
\begin{equation}\label{Eq Ep1}
\begin{aligned}
{E_{p}}&=2\left[\int_0^{\frac{S}{2}}\frac{{B{\kappa ^2}}}{2}ds+\int_0^{\frac{L}{2}}\frac{{C{\omega}^2}}{2}ds+\int_0^{\frac{S}{2}}w(\al,R)ds\right]\\
&=\int_0^{S/2}\left[B\ka^2+C\om^2+2w(\al,R)\right]ds+\frac{C\om_t^2(L-S)}{2},
\end{aligned}
\end{equation}
where, in the first line, the first two terms represent the bending and twisting energy respectively and the last term represents the inter-strand interaction.
Here \(B\) and \(C\) are the bending and torsional rigidities respectively, giving
\begin{equation}
B=\frac{1}{4}E\pi r^4,
\quad
C=\frac{1}{2}G\pi r^4,
\end{equation}
where $E$ is Young's modulus and $G$ is the shear modulus \cite{cable_DNA_writhe}. Likewise, the coefficient $2$ is given due to there being $2$ strands. For the purpose of generality, the expression of $w$, which depends on the specific problem, will be discussed later. However, at least, we can assume it to be a function of $\al$ and $R$ here.

We have assumed the formation process to be quasi-static and non-dissipative. Thus when the rod is in equilibrium, the energy minimization principle requires
\begin{equation}\label{energy minimization0}
\delta E_p= 0,
\end{equation}
which yields (See derivation in Appendix B)
\begin{equation}\label{eq dom/ds1}
\om=\left.\left(B\ka\frac{\pa\ka}{\pa\al}+\frac{\pa w}{\pa\al}\right)\right/C\frac{\pa\ta}{\pa\al},
\end{equation}
and
\begin{equation}\label{eq alR relation}
C\frac{\pa\ta}{\pa R}\left(B\ka\frac{\pa\ka}{\pa\al}+\frac{\pa w}{\pa\al}\right)-C\frac{\pa\ta}{\pa\al}\left(B\ka\frac{\pa\ka}{\pa R}+\frac{\pa w}{\pa R}\right)=0.
\end{equation}
It is noteworthy that this is a universal result, which is valid for both smooth and rough surfaces. Specifically, for the self-contact situation, we can assume $w(\al,R)$ to be $0$ for $R>r$ and infinity for $R<r$. In this case, according to \equref{eq alR relation}, it is possible only when $R=r$. Then, \equref{eq dom/ds1} can be simplified as
\begin{equation}\label{eq dom/ds1 self contact}
\om=\frac{B}{C}\cdot
\frac{\left(1-\cos2\al\right)\tan2\al}{2r}.
\end{equation}
\equref{eq dom/ds1 self contact} indicates that the range of $\al$ is confined to $0\leqslant\al\leqslant\pi/4$, since $\om$ would be negative if $\al>\pi/4$. The reason can be interpreted as follows. For $\pi/4<\al<\pi/2$, the curvature is still a monotonic increasing function of $\al$ whereas the torsion become a monotonic decreasing function of $\al$. However, the decrease of torsion will lead to the increase of twisting energy. Hence $\al$ must be less than $\pi/4$.
By applying two specific constraints to this variational problem, two results which correspond to the smooth and rough surfaces respectively will be investigated in the next subsections.

Assume the tension exerted by the rod is $T$ and the energy input from the environment is $W$. Then, the conservation of energy requires
\begin{equation}\label{force0}
dE_p=-TdS+dW,
\end{equation}
where $dW$ is the work done by twisting the rod,
\begin{equation}\label{eq dW}
dW=C\omega_t k(S)dS,
\end{equation}
where $k(S)$ is the twisting rate and it is defined by
\begin{equation}\label{k}
k(S)=2\pi\frac{ dN(S)}{dS}.
\end{equation}
It is noteworthy that $N(S)$ or $k(S)$ represents the specific process how the rod is twisted and slacked. 
\subsection{\label{sec:level2}Case \uppercase\expandafter{\romannumeral1}: Smooth Surface}
\begin{figure}
\begin{minipage}[t]{1\linewidth}
    \centering
    \includegraphics[width=0.5\textwidth]{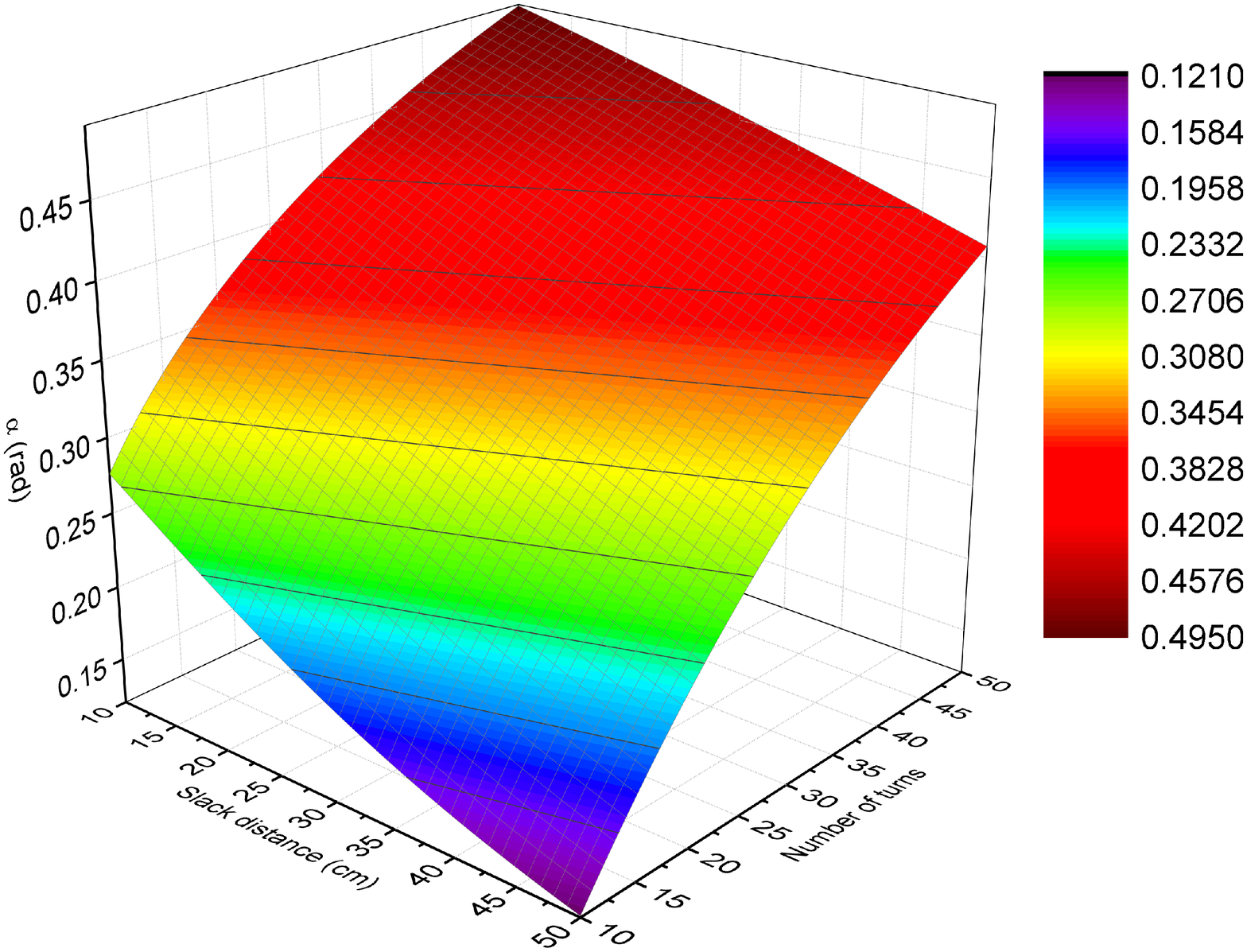}
\end{minipage}
\caption{Solutions of \equref{smooth surface}. Parameters used for the simulation: $r=1.0mm$, $L=57.0cm$, $E=4.20MPa$ and $G=1.51MPa$.}\label{figure2}
\end{figure}

Here we consider the situation in which the rod has no surface friction, namely, $f=0$. Therefore no external twist moment is exerted on the rod. Then \equref{eq twist moment der} and \equref{eq twist moment} yield, for $s<S/2$,
\begin{equation}\label{eq dw/ds1}
\frac{d \om}{d s}=0.
\end{equation}
Since \equref{eq dom/ds1} indicates $\al$ depends only on $\om$, \equref{eq dw/ds1} can yield
\begin{equation}
\frac{d\al}{ds}=0.
\end{equation}
That means, for a rod without surface friction, the double helix formed is uniform. Note here the influence of the head and the conjunction between the tail and helix is neglected. Then by using the Calugareanu invariant \equref{Lk=Tw+Wr} and \equref{Eq Lk=Tw+Wr}, $\om$ can be expressed in terms of $\tau$,
\begin{equation}\label{smooth surface}
\om=\frac{2\pi N-\ta S}{L}
\end{equation}
By solving this equation, we can determine the configuration of the double helix. One typical solution for the self-contact situation is shown in \reffig{figure2}. Since there is no surface friction, the configuration of the double helix only depends on $N$ and $S$. It is independent of the twisting and slacking history.

\subsection{\label{sec:level2}Case \uppercase\expandafter{\romannumeral2}: Rough Surface}
In this case, we will consider the situation in which the two strands are rough and have a surface friction. The surface friction is more common for classical models where the two strands are self-contacted. Therefore, here we can assume $w$ to be $0$ for $R>r$ and infinite for $R<r$. As noted above, \equref{eq dom/ds1 self contact} can be obtained from this assumption. The surface friction will contribute to the twist moment, as shown in \equref{eq twist moment}. Thus $d\om/ds$ can be nonzero for the double helices formed by rods with rough surface. According to \equref{eq dom/ds1}, nonzero $d\om/ds$ means that the double helix is non-uniform.
Here, we consider a typical situation in which the surface is rough enough so that there is no relative slipping between the two strands. That means, the established double helix will have fixed spatial configuration and twist, and it will not be influenced by the followed newly formed double helix. Thus, for the helix part $s<S/2$, there is no S dependence for $\al$
\begin{equation}
\frac{\partial \al}{\partial S}=0.
\end{equation}
Since there is no explicit S-dependence in the expression of $\al$, we can obtain (See all the derivation for this subsection in Appendix C)
\begin{equation}\label{eq da/ds}
\frac{d\al}{ds}=\frac{C}{B(L-2s)}\cdot\frac{\left[\sin2\al-2Rk(2s)\right]\cos^22\al}{\cos^32\al -1},
\end{equation}
The corresponding initial condition is determined by
\begin{equation}\label{eq ini}
\left.(1-\cos2\al)\tan2\al\right|_{s=0}=\frac{C}{B}\cdot\frac{4\pi RN}{L}.
\end{equation}
 By solving this equation, the spatial configuration of the double helix can be determined. Then $P$, $\tau$ and $\kappa$ can be obtained by using \equref{eq alpha def} \equref{Eq curvature1} and \equref{Eq torsion1}.
 We can obtain the tension exerted by the rod by using \equref{force0} as well
\begin{equation}\label{eq tension}
T=\left.\frac{B}{8R^2}\cdot\frac{(1-\cos2\al)^2(2+\cos2\al)}{\cos2\al}\right|_{s=S/2}.
\end{equation}

\subsection{\label{sec:level2}Surface friction}
In case II, we consider the rough surface and assume that the surface friction is large enough so that there is no relative slipping between the two strands. In this section we will investigate the surface friction quantitatively and examine the validity of the assumption for non-relative slipping. To obtain the surface friction, we must first know the normal reaction between the two strands. It can be estimated as follows.
We assume the radius $R$ is not a fixed parameter here. For double helix in equilibrium, the configuration can be regarded as a function of $R$. Consider a piece of double helix with an infinitesimal length $\delta S$ in the whole double helix. Since $\delta S$ is infinitesimal, the double helix can be considered to be uniform approximately. This piece of double helix is assumed to be isolated from the rest of the double helix. Therefore the Calugareanu invariant can be applied to it. By considering the relation between R and the potential energy we can be obtain the normal reaction per unit length (See the derivation in Appendix D)
\begin{equation}\label{eq normal reaction}
F=\frac{B}{8R^3}\cdot\frac{(1-\cos2\al)^2}{\cos2\al}.
\end{equation}
Therefore, the maximum static friction for unit length is approximately $f_{max}=\mu F$, where $\mu$ is the coefficient of friction. Then by using \equref{eq twist moment der}, \equref{eq twist moment} and \equref{eq dM_f}, we can obtain the condition for no relative slipping
\begin {equation}\label{eq friction con}
C\left|\frac{d\om}{ds}\right|\leqslant \frac{dM_{fmax}}{ds},
\end{equation}
where $dM_{fmax}/ds$ is defined by
\begin{equation}\label{eq friction}
\frac{dM_{fmax}}{ds}=\frac{B\mu}{8R^3}\cdot\frac{\cos\al(1-\cos2\al)^2}{\cos2\al}.
\end{equation}
This result indicates that only when $d\om/ds$ is less than a critical value, the external twist moment provided by the surface friction can guarantee no relative slipping. The simulation in the next section indicates that this condition can be satisfied well in most cases.

\section{\label{sec:level1}Applications}
By using this model, the formation of non-uniform double helices for rods with rough surfaces can be analyzed and predicted. We can even produce a double helix with a designed configuration by controlling the twisting and the slacking process. For illustration, here we consider three typical examples to show the power of this model.
\subsection{\label{sec:level2}Example I}
One typical situation is that the rod is initially twisted by $N$ turns and then the two ends are got together without increasing $N$, namely, $k=0$.
It is the most common situation in both everyday life and research. In the CNT yarn twisting experiments performed by Cao \textit{et al.} \cite{CNT1}, similar procedure was used and thus non-uniform double helices were produced. Here we try to investigate this phenomenon.
The solution for \equref{eq da/ds} with $k=0$ is
\begin{equation}\label{eq example k=0}
\frac{\cos2\al+1}{\cos2\al_0+1}e^{\sec2\al_0-\sec2\al}=\left(1-\frac{2s}{L}\right)^{-\frac{C}{B}},
\end{equation}
where $\al_0$ is the initial value of $\al$, which can be determined by \equref{eq ini}. From \equref{eq example k=0}, it can be found the double helix will become increasingly looser with the increase of $s$. Besides, the pitch variation near the head is much slower than that near the tail (shown in \reffig{kzero:subfig}). Therefore, in order to obtain relatively uniform double helices, for example fabricating the CNT superconductors, we should try to avoid this method or, at least, only focus on the double helix near the head. The comparison between the experiment results and the simulations are shown in the next section.
\subsection{\label{sec:level2}Example II}
According to \equref{eq da/ds}, it can be found that we may control the configuration of the double helix by manipulating the twisting and slacking process, namely, controlling $k(S)$, which is defined in \equref{k}. Due to the similarity between the classical elastic rods and the CNT yarns, this technique can be used to control the configuration of the double helix of CNT yarns during its fabrication. Here for the purpose of demonstration, we adopt the silicone rubber rod and choose the pitch distribution in this form
\begin{equation}\label{eq sinp}
P(s)=P_0+P_1\sin(ks).
\end{equation}
When the rod is twisted and slacked, there are three variables that we can control directly, namely, $N$, $S$ and $T$. Among these three variables, only two are independent. Therefore we can obtain the expected double helix by controlling only two of them simultaneously. By using \equref{eq alpha def}, \equref{eq da/ds} and \equref{eq tension}, we can obtain the relation for $N$, $S$ and $T$ easily. However, it will be difficult to exert a tension as a function of $S$ or $N$ experimentally. Therefore it is more feasible to control $N$ and $S$ simultaneously. More details about the experiments are shown in the next section.

\subsection{\label{sec:level2}Example III}
The third one is to produce a uniform double helix with fixed $P$ and $\al$ and use this model to analyze the DNA supercoiling experiment performed by Strick \textit{et al.} \cite{twisting_DNA2,Strick1998}. Here, we will  consider the smooth surface situation and the reason will be discussed later. \equref{eq dom/ds1} indicates that fixed $P$ and $\al$ require fixed $\om$, namely, $\pa\om/\pa S=0$. Then by differentiating both sides of \equref{smooth surface} with respect of $S$, we can obtain the twisting rate $k$,
\begin{equation}\label{eq kappa DNA}
k=\ta.
\end{equation}
This equation indicates that, during the formation process, the twisting is completely transformed into the spatial writing of the DNA. The tension can be easily obtained by using \equref{Eq Ep1}, \equref{force0} and \equref{eq kappa DNA},
\begin{equation}
T=-\frac{1}{2}B\ka^2+C\om\ta-w.
\end{equation}
Some articles used the roughly estimated critical tension $T=M^2_3/2B$ for buckling transition as the tension for the supercoiling formation \cite{Strick2003,self_contact1,Salerno2012}. This approximation is oversimplified, since the configuration changes for the loop formation in the buckling and the double helix extension are completely different. Besides, the buckling transition is a sudden process where the tension is extremely unstable. Thus, it is more reasonable to consider the tension for double helix extension directly.

Due to the thermal fluctuation of the DNA chains, there will be a repulsion between the two strands of the double helix. Besides, each segment of the polymer chains will take up a certain volume and they cannot overlap with each other. To describe these statistical properties of the DNA chains, here we follow the method of Marko and Siggia \cite{Marco1994}, for DNA supercoiling, $w(\al,R)$ can be written as,
\begin{equation}
\begin{aligned}
w=&\frac{\left[ 1+(\pi\cot\al)^{-12}\right] k_Bt}{r}\cdot(R/r)^{-12}\\
&+\frac{B^{-1/3}[1+(\pi \cot\al)^{-2/3}](k_Bt)^{4/3}}{R^{2/3}},
\end{aligned}
\end{equation}
where the first term represents the hard core interaction and second terms represent the entropic loss for winding too tightly. To avoid ambiguity, we use $t$ to represent the temperature here. Here is the reason why the smooth surface model is used. Since the tension is fixed, the double helix formed must be uniform. This result is still valid even if we use the rough surface model developed above. For the uniform double helix, \equref{eq twist moment der} and \equref{eq twist moment} indicate that the surface friction must be $0$. Thus, if a fixed tension is applied, it does not matter whether the surface is rough or not. Both models will yield the same results.

\begin{figure}
  \centering
\begin{minipage}[t]{1\linewidth}
    \centering
    \includegraphics[width=0.5\textwidth]{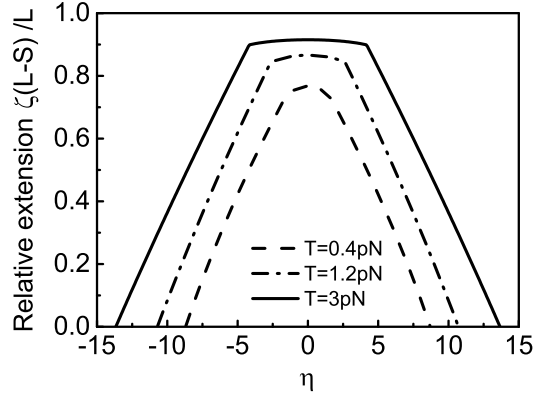}
\end{minipage}
  \caption{Simulation for the DNA twisting experiment. Tension (corresponding $R$ calculated): 0.4pN (2.493nm), 1.2pN (2.299nm) and 3.0pN (2.134nm). $\eta=2\pi N \xi_T/L$ represents the degree of supercoiling. Note here the thermal fluctuation and the loop formation at the head are ignored. }\label{DNA}
\end{figure}

The thermal fluctuation will cause the tail to deviate from the straight line and thus to tangle spatially.
In this case, the distance between the two ends will be less than the real length of the tail. Moreover, $Lk$ will be partly stored in the form of $Wr$ for the tail part.
To take these factors into account, we can utilize the method proposed by Moroz and Nelson \cite{Moroz1997}. The effective torsional rigidity $C_{eff}$ for the chain with thermal fluctuation can be expressed as
\begin{equation}
\frac{1}{C_{eff}}=\frac{1}{C}+\frac{k_Bt}{4B\sqrt{BT}}.
\end{equation}
Due to the restriction of \equref{eq twist moment der}, $\om$ for the tail part should be multiplied by a factor $C/C_{eff}$ here.
The ratio between the distance of the two ends and the real length of the tail can be expressed as
\begin{equation}
\begin{aligned}
\zeta=&1-\frac{1}{2}\left({\frac{BT}{k_B^2t^2}-\frac{M_3^2}{4k_B^2t^2}-\frac{1}{32}}\right)^{-1/2}\\
&+\frac{Bk_Bt}{(L-S)\left(BT-\frac{M_3^2}{4}\right)}.\\
\end{aligned}
\end{equation}
Therefore the distance between the two ends should be $\zeta(L-S)$.
Combined with these modifications, the relation between the end-to-end distance and the number of turns can be calculated. The simulation is shown in \reffig{DNA}, where, for clarity, the degree of supercoiling $\eta=2\pi N \xi_T/L$ is used instead of $N$. Here $\xi_{T}$ is the persistence length used in the worm-like-chain model of DNA.
The parameters used for the simulation are listed as follows, bending modulus $B/k_BT=48nm$, torsional modulus $C/k_BT=86nm$ \cite{DNAtorsion}, $T=300K$, $r_0=1.75nm$ \cite{Vologodskii1992}, DNA length $L=750nm$ and persistent length $\xi_{T}=52nm$ \cite{chainmodel}. Note here the possible denaturation transition of the DNA chains is ignored, which causes the symmetric graph of \reffig{DNA}. For tension over than $0.8pN$, the negative supercoiling may have denaturation \cite{Salerno2012}. Therefore our model cannot used for the situation of negative supercoiling with large tension.

\section{\label{sec:level1}Experiment Results}
\begin{figure}
\subfigure[]{
    \includegraphics[width=0.5\linewidth]{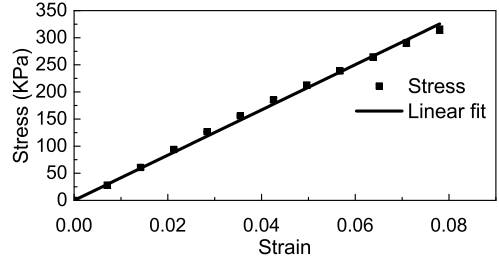}
    \label{modulus:subfig:1}
}
\subfigure[]{
    \includegraphics[width=0.5\linewidth]{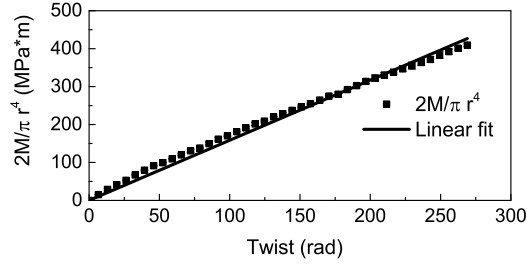}
    \label{modulus:subfig:2}
}
\caption{Measurement of the Young's and shear modulus for the silicone rubber rod. (a) Tensile test result. Young's modulus measured: 4.20MPa. (b) Twisting test result. Shear modulus measured: 1.51MPa.}\label{modulus:subfig}
\end{figure}

Our experiment uses silicone rubber rods with circular cross section provided by Shanghao Daoguan Rubber and Hardware Co. Ltd. Its chemical formula is
$[-R_2SiO-]_n$ with $R$ representing methyl group or phenyl group. Its hardness is $35-75$ shore A, according to the provider. The Young's and shear modulus of the rubber rod are measured by tensile and twisting tests. The results are shown in \reffig{modulus:subfig}, by which we can obtain $E=4.20MPa$ and $G=1.51MPa$. It is noteworthy that when the strain exceeds 0.08, the rod has an obvious nonlinear behavior. Since in our following experiments for the rod, the strain is kept less than 0.08, here we only consider and fit the data points with strain less than 0.08. The possion's ratio is 0.4, which is close to the value 0.44 in literature \cite{possion2}.
\par
\begin{figure*}
\subfigure[]{
    \includegraphics[width=0.48\linewidth]{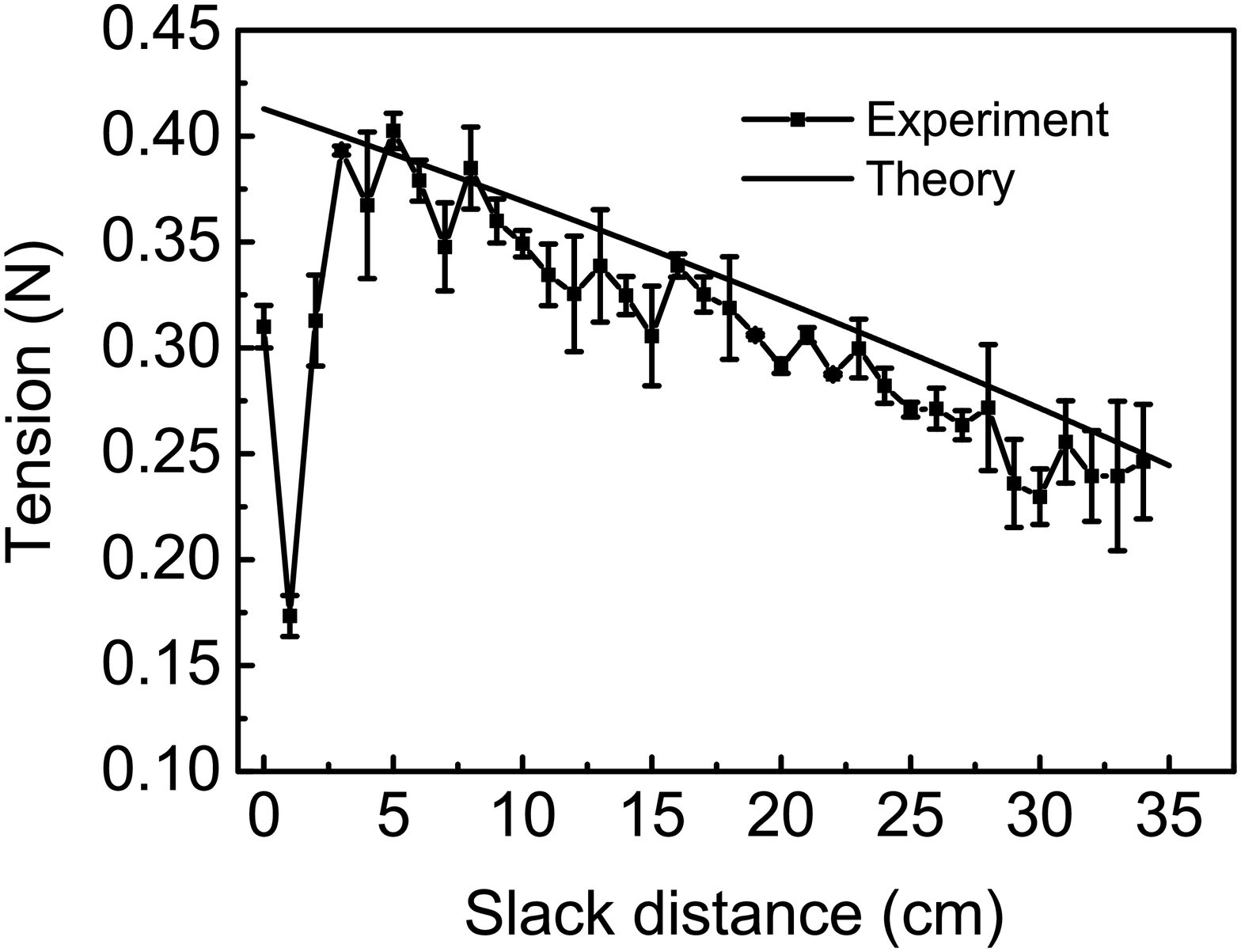}
    \label{tt:subfig:1}
}
\subfigure[]{
    \includegraphics[width=0.48\linewidth]{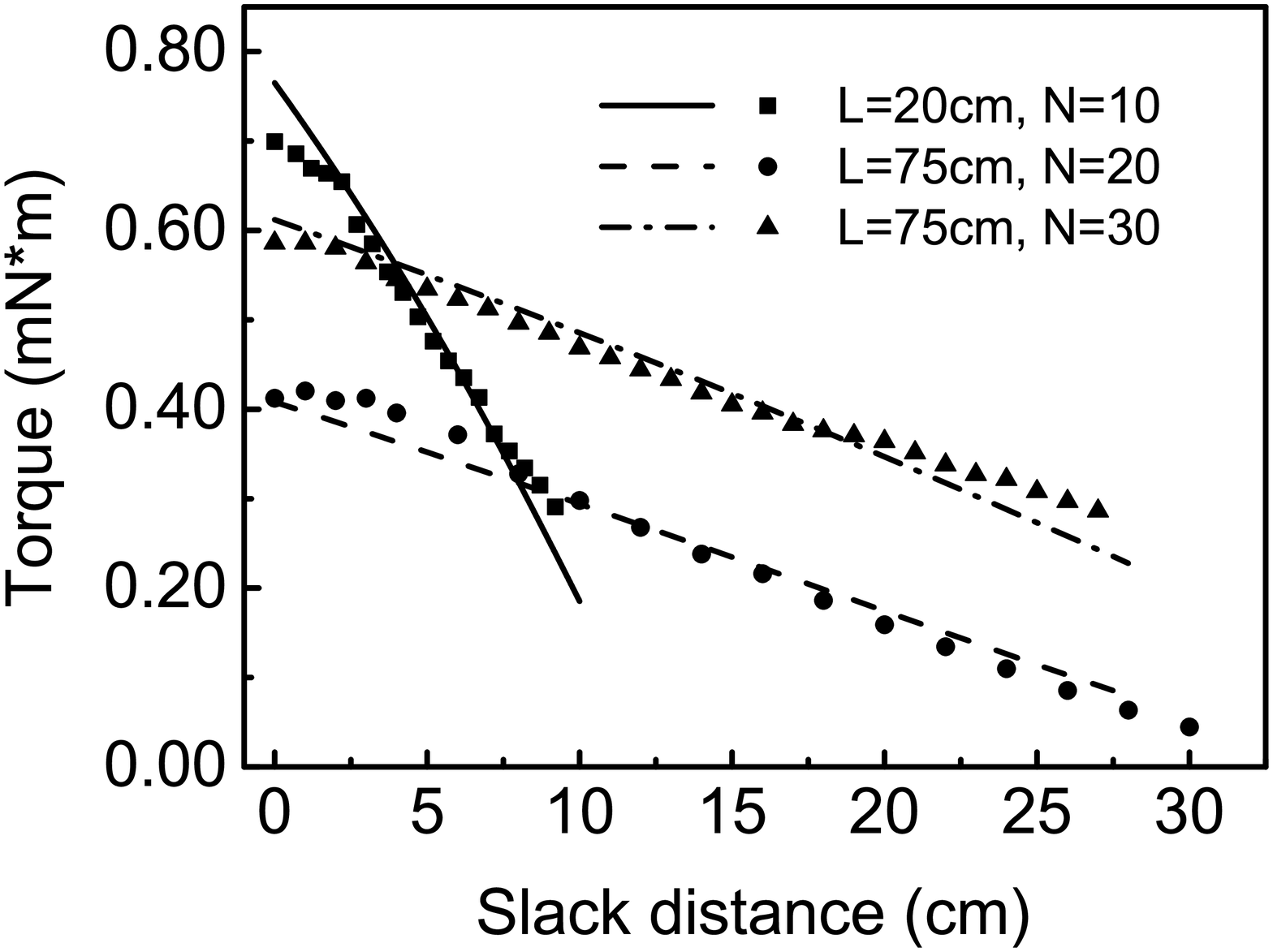}
    \label{tt:subfig:2}
}
    \caption{Tension and torque exerted by the twisted rod when it is slacked. Dots: experiment data. Curves: theoretical predication. (a) The tension measured at the fixed end. Note there is a sudden decrease of $T$ at $S\sim1cm$ due to the jump in localized buckling. Rod parameters: r=1mm, L=1.01m, N=80. (b) The torque measured at the fixed end. Rod parameters: $r=1.0mm$.}\label{tt:subfig}
\end{figure*}
\begin{figure*}
\subfigure[]{
    \includegraphics[width=0.48\linewidth]{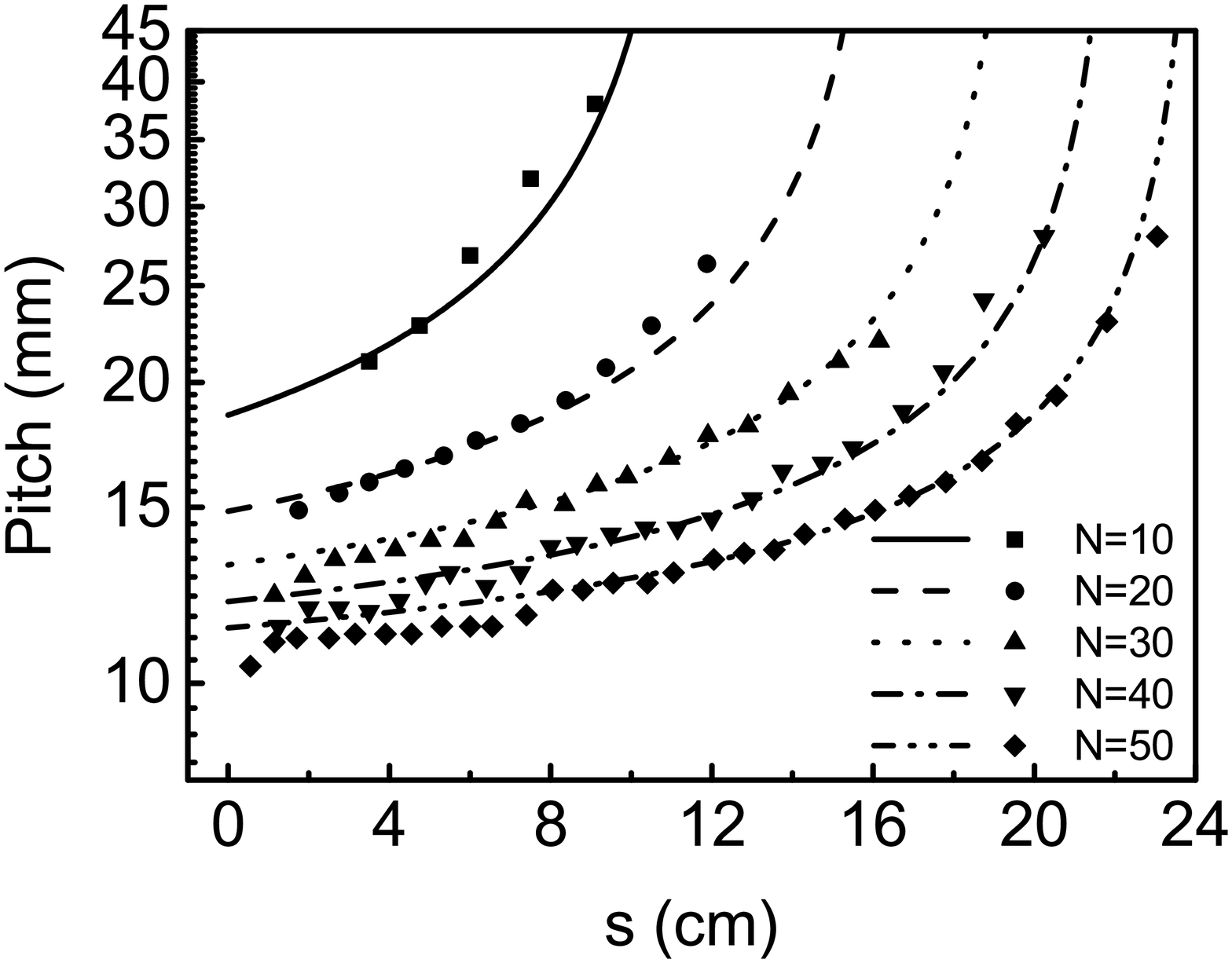}
    \label{kzero:subfig:1}
}
\subfigure[]{
    \includegraphics[width=0.48\linewidth]{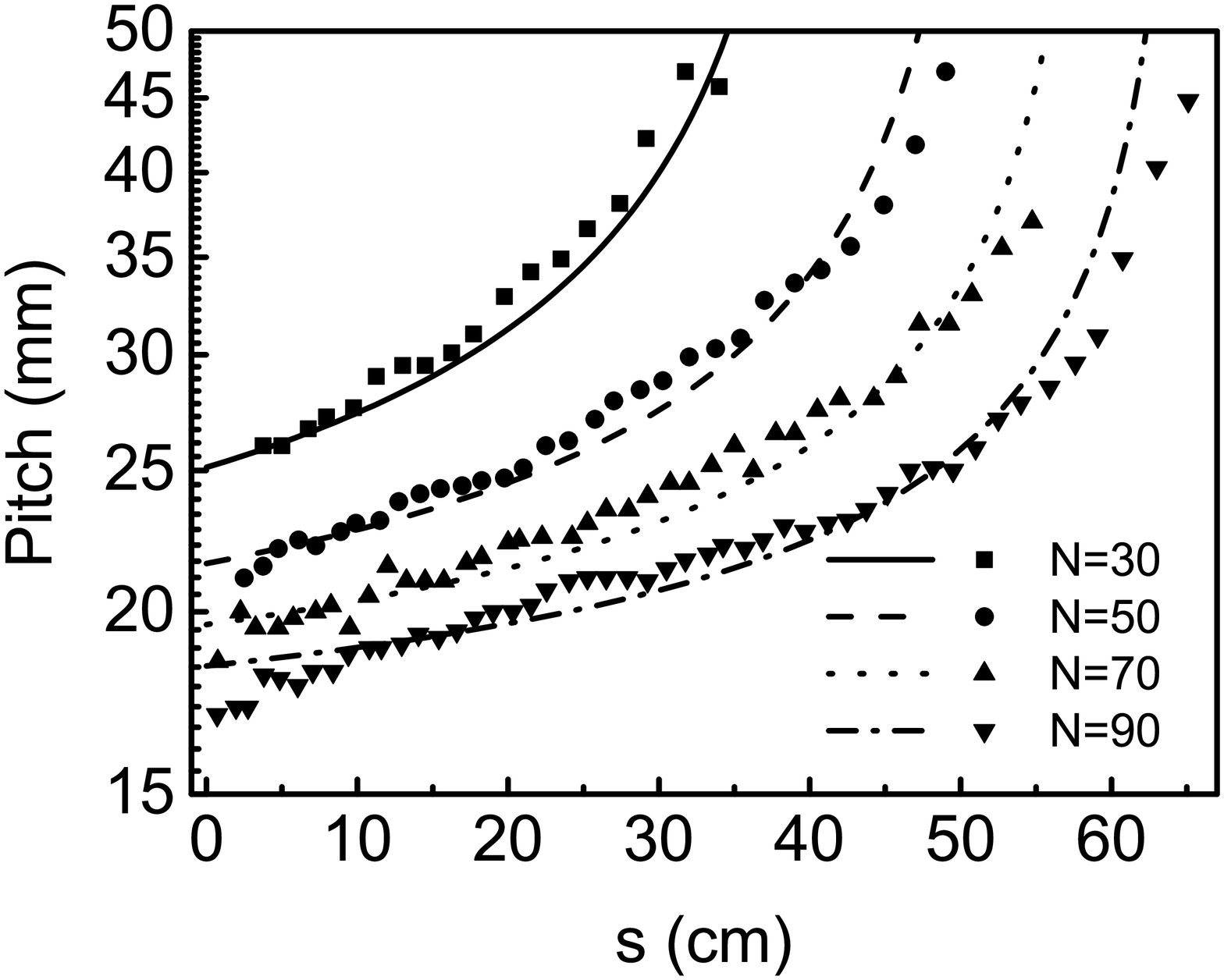}
    \label{kzero:subfig:2}
}

    \caption{Pitch variation for the first example in which the rod is initially twisted by $N$ turns and then slacked without being twisted (k=0). Curves: theoretical predications. Dots: experiment data. The x-axis represents the natural coordinate $s$. Rod parameters: (a) $r=1.0mm$, $L=56.7cm$. (b) $r=1.7mm$, $L=142.0cm$}\label{kzero:subfig}
\end{figure*}
\begin{figure*}
  \centering
  \includegraphics[width=0.9\textwidth]{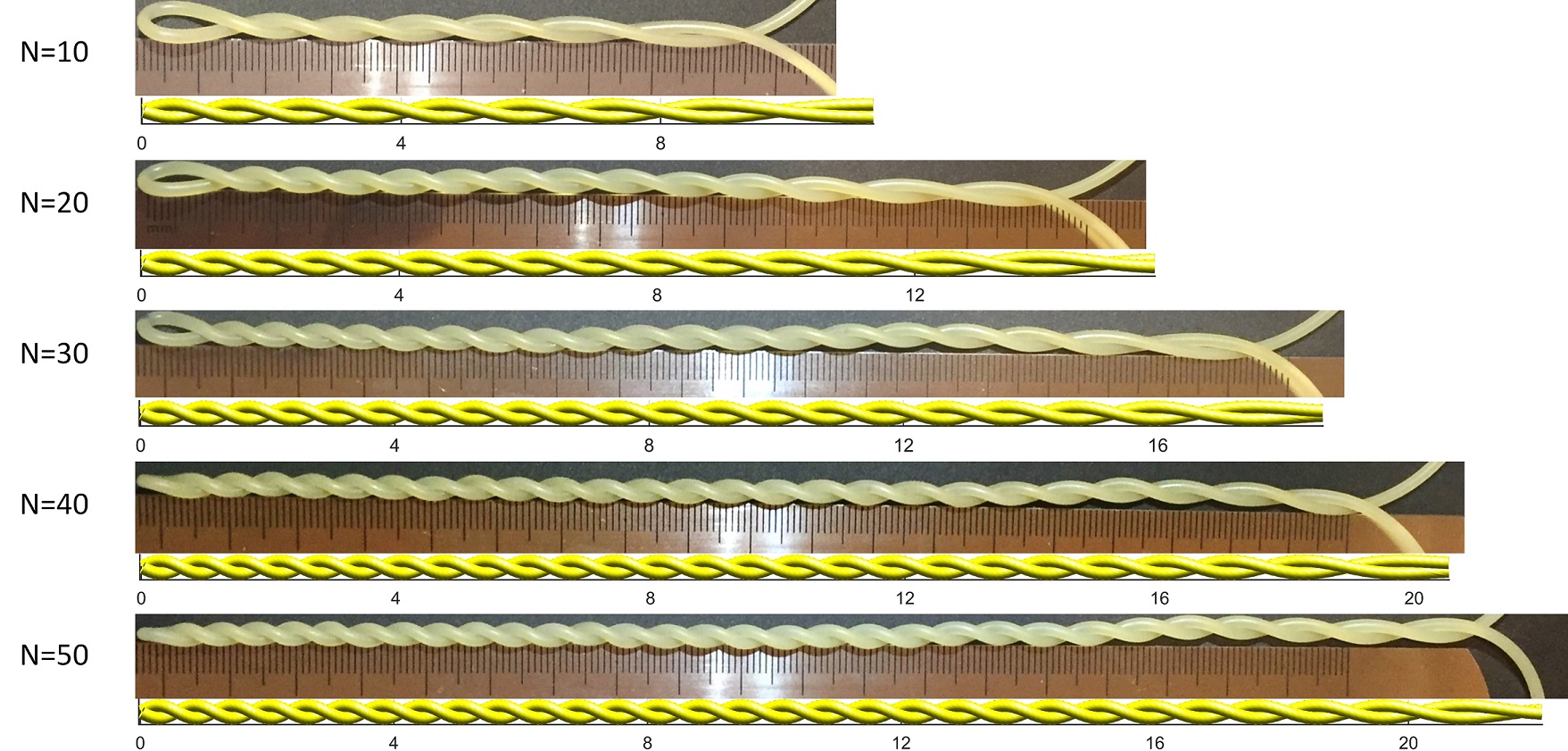}
  \caption{The non-uniform double helices obtained by the experiments for the first example. The simulation results are plotted with the same length scale for comparison. Rod parameters: L=57cm and r=1mm.}\label{real1}
\end{figure*}

For the first example mentioned in Case II. The rod is initially twisted by $N$ turns. One end is fixed while the other end is moveable. The main part of the rod is placed on a smooth flat surface. The height of its two ends is controlled to be less than 0.5cm to ensure that the tension induced by the gravity is as small as possible $(0.01\sim0.02N)$. The slider is slowly moved and the double helix is formed at the mid-point of the rod. 
The tension and torque are measured at the fixed end respectively. 
the results are shown in \reffig{tt:subfig:1}. Note here rods with different lengths are used for measuring the tension and torque due to the restriction of the instrument. In \reffig{tt:subfig:1}, it is noteworthy that a sudden decrease of the tension occurs at $S\sim 1cm$. It is due to the jump in localized buckling which is accompanied by the formation of the first loop at the head. This formation process has been discussed broadly \cite{experiment1}.
After the formation of the whole double helix, the pitch distribution is measured. The results are shown in \reffig{kzero:subfig}. For comparison, the theoretical predications are plotted as well. 
Some of the non-uniform double helices obtained are shown in \reffig{real1}, where the corresponding simulation results are plotted for comparison. It can be seen that the pitch increases from the head to the tails in both figures.
In all the experiments mentioned above, we did not observe relative slipping for the two strands of the double helix. This provides a verification for the assumption of Case II, which states that the two strands are fixed after the double helix is formed.
 The lengthening induced by the twisting for our rod is less than $2\%$, which guarantees the inextensible assumption for our model.

\begin{figure*}
\subfigure[]{
    \includegraphics[width=0.32\linewidth]{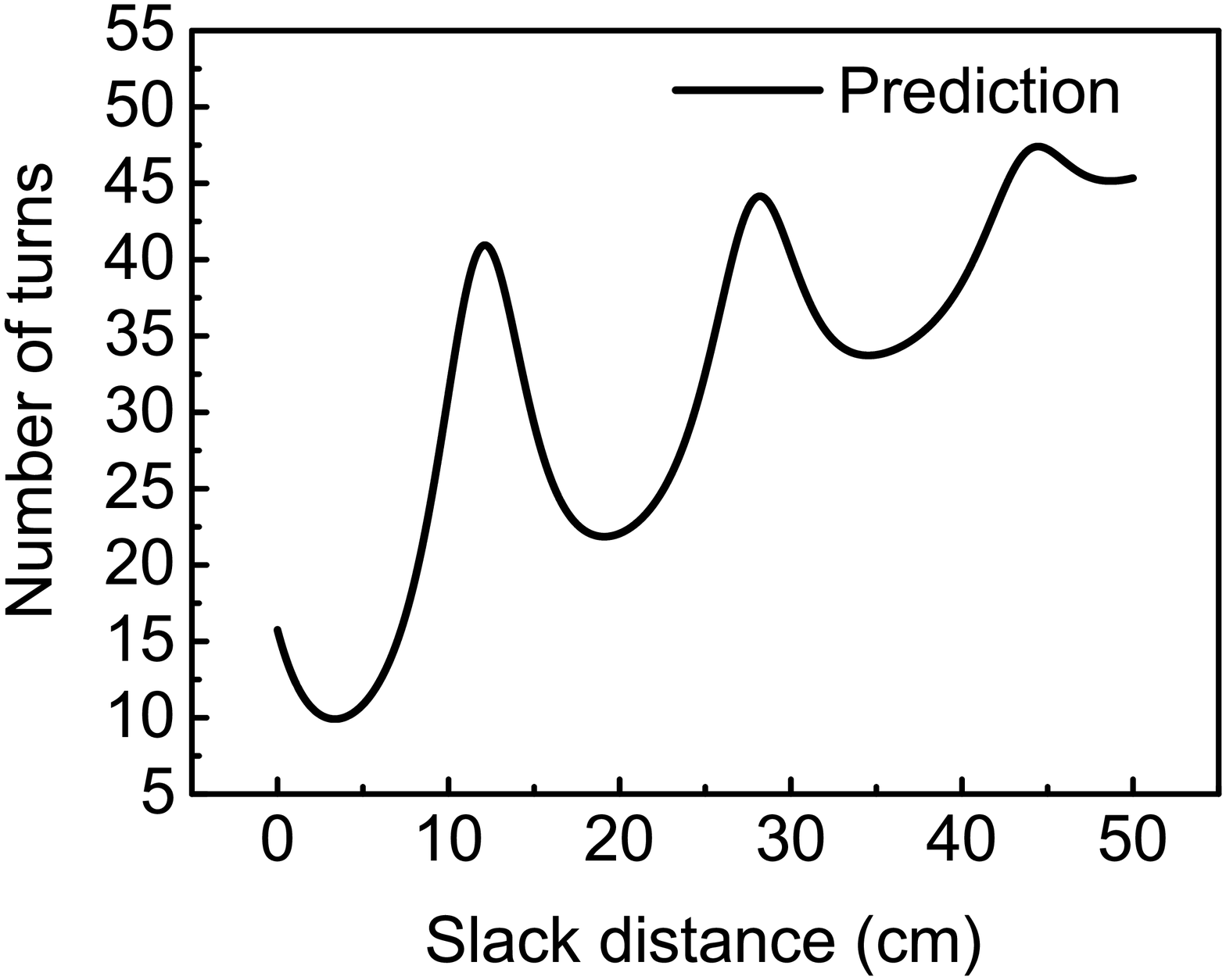}
    \label{sin:subfig:1}
}
\subfigure[]{
    \includegraphics[width=0.32\linewidth]{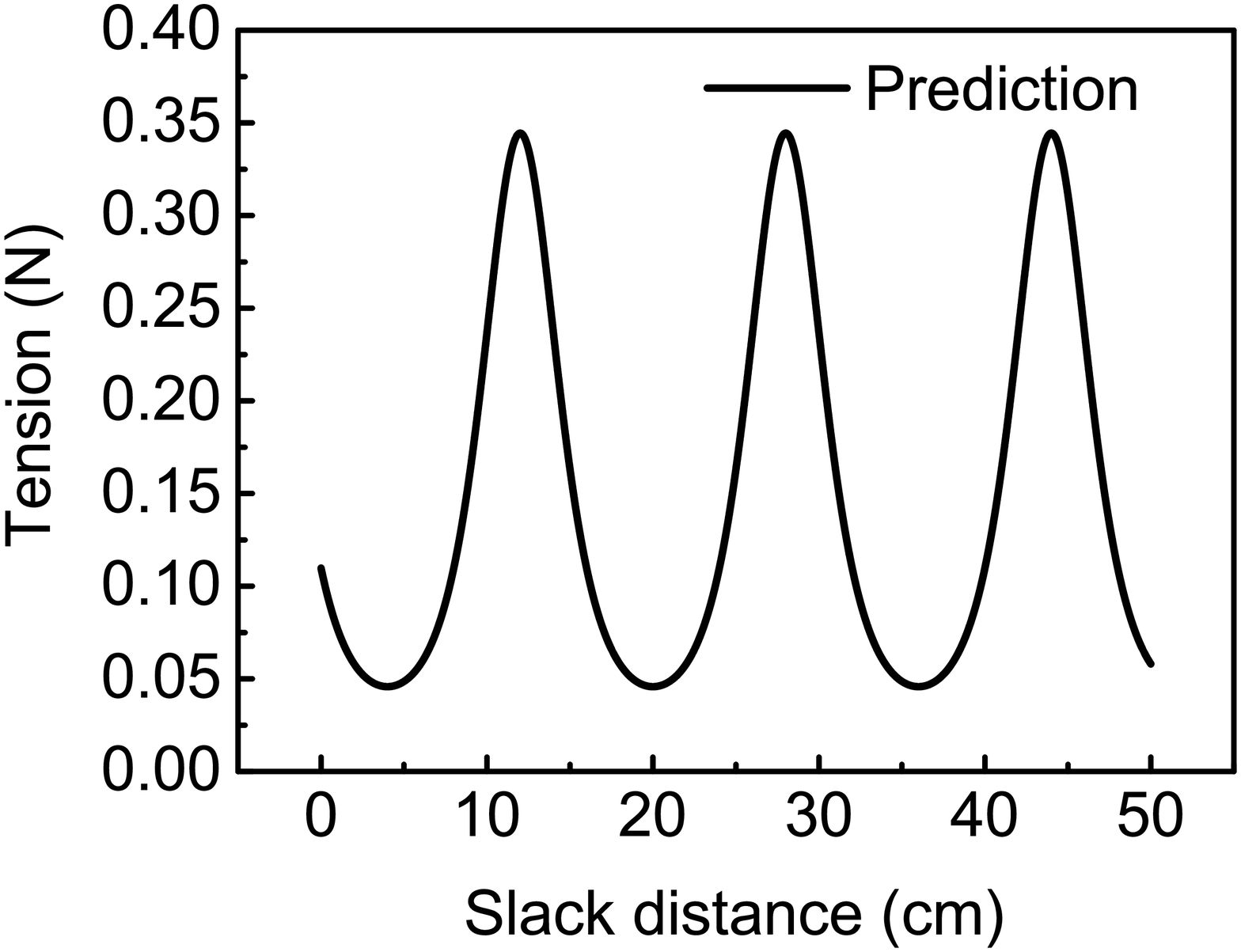}
    \label{sin:subfig:2}
}
\subfigure[]{
    \includegraphics[width=0.32\linewidth]{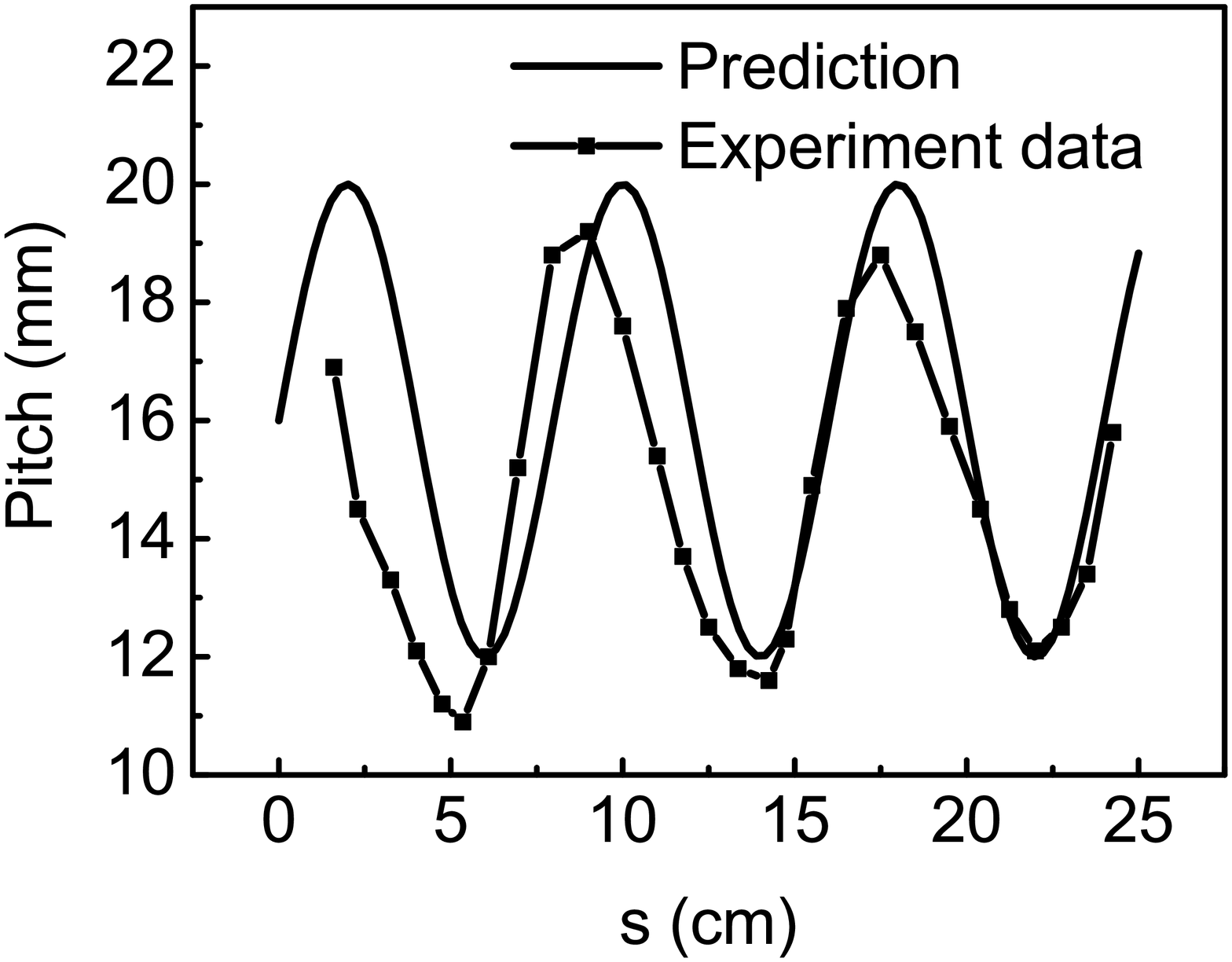}
    \label{sin:subfig:3}
}
\caption{Producing a double helix with a wavelike pitch distribution: $P(s)=P_0+P_1\sin(ks)$, where $P_0=16cm$, $P_1=4cm$ and $k=2\pi/10P_0$. (a) Expected relation between the slack distance and the number of turns. (b) Expected relation between the tension and the number of turns. (c) Pitch distribution. Solid line: theoretical prediction. Dots: experiment data.}\label{sin:subfig}
\end{figure*}

\begin{figure*}
  \centering
  \includegraphics[width=0.9\textwidth]{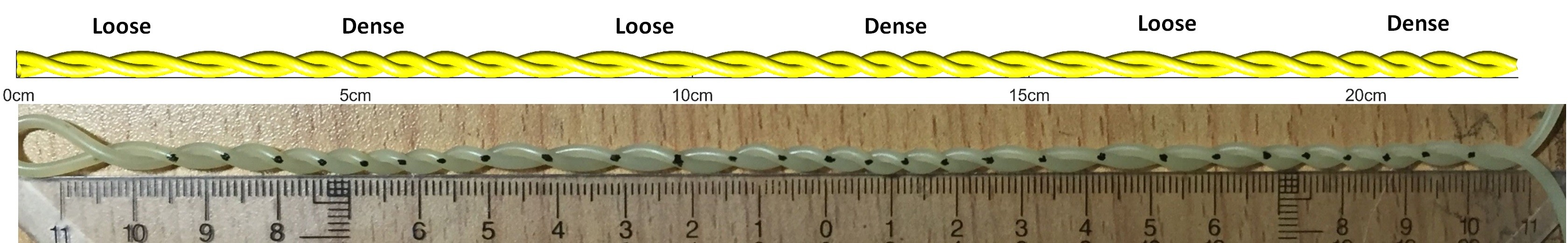}
  \caption{The experimentally obtained double helix with a wave-like pitch distribution for the second example. The simulation result is plotted with the same length scale for comparison. Rod parameters: L=57cm and r=1.0mm.}\label{real2}
\end{figure*}

For the second example mentioned in Case II, we choose $P_0=16mm$, $P_1=4mm$ and $k=2\pi/10P_0$. The predicted relation for $N$, $T$ and $T$ is plotted in \reffig{sin:subfig}(a,b). Here we choose to control $N(S)$ to obtain the designed double helix for the reason mentioned previously. The double helix obtained is shown in \reffig{real2}, of which the pitch distribution is shown in \reffig{sin:subfig:3} and compared with the theoretical predication. It is noteworthy that the pitch near the head is slightly less than theoretically expected value due to the influence of the loop at the head.

\begin{figure*}
\subfigure[]{
    \includegraphics[width=0.48\linewidth]{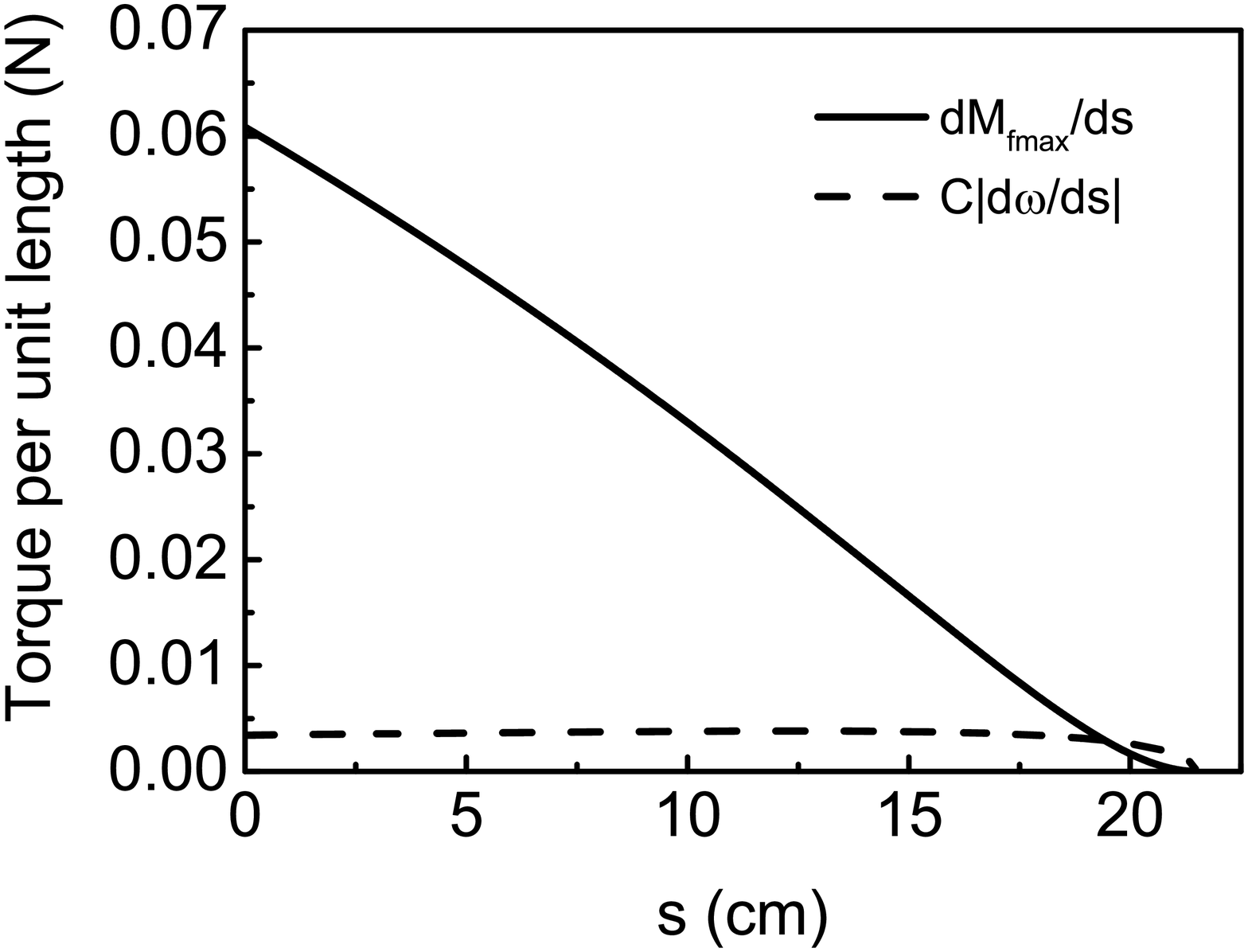}
    \label{fricth:subfig:1}
}
\subfigure[]{
    \includegraphics[width=0.48\linewidth]{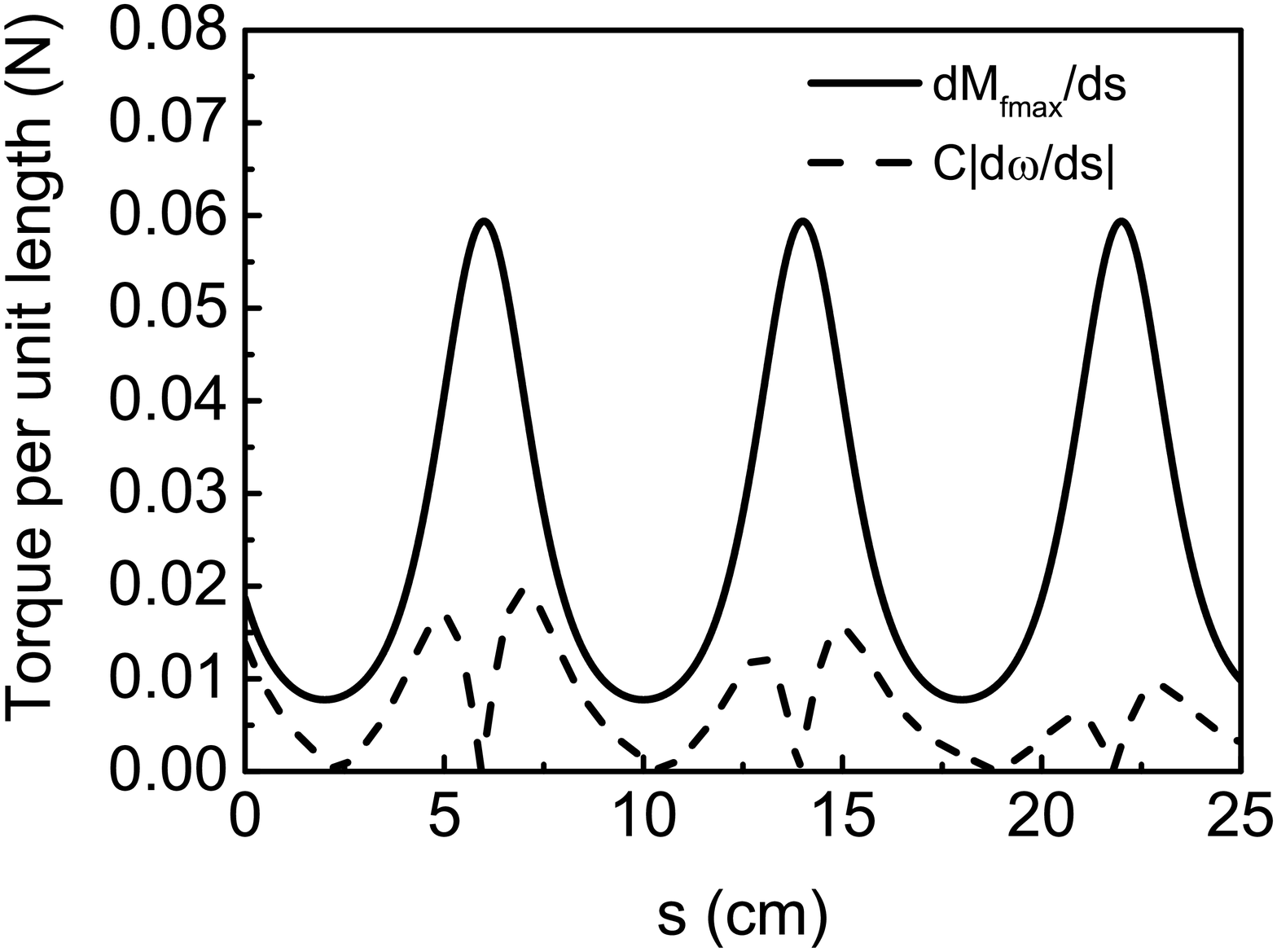}
    \label{fricth:subfig:2}
}

\caption{Theoretically calculated $C|d\omega/ds|$ and $dM_{fmax}/ds$ variations for (a) the N=40 situation in \reffig{kzero:subfig:1}, and (b) the artificially designed double helix in \reffig{sin:subfig:2}.}\label{fricth:subfig}
\end{figure*}

\section{\label{sec:level1}Discussion}
\subsection{\label{sec:level2} Maximum helix length}
In the experiments for the first example, it can be found that pitch diverges to infinity before the two ends reach each other. In other words, the formation of the double helix is finished before $S$ reaches $L$. The explanation is as follows. From \equref{eq example k=0}, It can be found that the range of the left-hand side is confined to $0\sim2/e$ whereas the right-hand side will tend to infinity when $s\rightarrow S/2$.
Hence, the maximum value of $s$ for the helix part must be less than $L/2$. We denote the maximum value by $s_{max}$ here. It can be calculated from \equref{eq example k=0} by letting $\al$ be 0,
\begin{equation}
s_{max}=\frac{1}{2}\left[1-\left(\frac{2e^{\sec2\al_0-1}}{\cos2\al_0+1}\right)^{-\frac{B}{C}}\right]L.
\end{equation}
However this restriction on the length of the double helix is not complete. Note that this model requires the surface friction to be large enough to prevent the relative slipping. When $s \rightarrow s_{max}$, the double helix becomes very loose. Then the condition \equref{eq friction con} for no relative slipping is not satisfied anymore. This factor will slightly influence the configuration of the double helix. For illustration, we calculate $C|d\om/ds|$ and $dM_{fmax}/ds$ for the $N=40$ situation in \reffig{kzero:subfig:1}. Note the coefficient of friction $\mu$ for the silicone rubber rods is taken to be 0.5. See the method of measuring $\mu$ in Appendix E. The result is shown in \reffig{fricth:subfig:1}. It can be found that \equref{eq friction con} can be satisfied unless $s$ is extremely close to $s_{max}$. Therefore the non-relatives-slipping assumption is valid for most part of the double helices. As a matter of fact, the experiments show that the double helices are slightly longer than the theoretical predictions due to the relative slipping for $s \rightarrow s_{max}$. As for the second experiment about producing a double helix with a designed configuration, we calculated its $C|d\om/ds|$ and $dM_{fmax}/ds$ as well. The result is shown in \reffig {fricth:subfig:2}. It can be found that \equref{eq friction con} can be satisfied well.
\subsection{\label{sec:level2} Main advantages}
Compared with the conventional twist-spinning method to produce double helix \cite{gao2016,braiding_DNA1}, this twist-slacking process has unique advantages.
In the previous twist-spinning method, the two strands are spun individually and self-twisted to form double helix. Simultaneously, a tension is exerted on the ends to keep the double helix straight. In the twist-slacking method used here, the rod is only twisted and pulled at the ends. 
One advantage of the the twist-slacking method is that its manipulation is much simpler than that of the twist-spinning method, which accounts for the fact that, the former is more common than the latter, especially in microscopic regime. As noted previously, both the classical DNA-twsiting experiments \cite{twisting_DNA2} and the double helix formation of CNT yarns used this method. Its simplicity originates from the fact that we only need to hold and twist one end of the rod while the other end is keep fixed. However, for the twist-spinning process, we need to twist both ends simultaneously and perform a tension to the double helix. Although it can be easily realized for rods with macroscopic scale, for example, thick CNT yarns and wires. Nevertheless, there is a formidable difficulty in performing all these steps simultaneously for microscopic rods. For example, it is hard to spin the two strands simultaneously. In order to braid DNA chains, Croquette used the magnetic bead to attach two DNA chains and twist them as a whole directly without spinning the two individual strands \cite{braiding_DNA1}. In this case, the internal twist of the double helix will accumulate, which may eventually result in the supercoiling. 

\begin{figure}
  \centering
\begin{minipage}[t]{0.7\linewidth}
    \centering
    \includegraphics[width=0.5\textwidth]{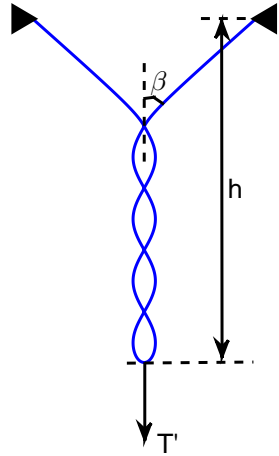}
\end{minipage}
  \caption{Schematic of the double helix obtained by the twist-spinning method.}\label{withforce}
\end{figure}

Besides its simplicity, another advantage is that the double helix obtained by the twist-slacking method is in equilibrium state automatically once it is formed. In contrast, it is hardly for the twisting-spinning method to achieve these. Due to the influence of the tension to keep the double helix straight, the pitch of the double helix will be enlarged. After the formation process, the tension should be removed. In this case, the double helix will reach a new equilibrium state and the pitch will increase automatically, which means the configuration of the double helix will have a deviation. To illustrate this and evaluate the degree of the deviation, we consider a simple model involving the twist-spinning process here. Now assume a tension $T'$ is exerted on the head of the double helix while the head can still rotate freely without any restriction, as illustrated in \reffig{withforce}. Obviously the angle between the two tails, denoted by $2\be$ here, will lessen due to the existence of $T'$.
All the other procedures are the same as before. The length of the rods composing the helix part is $S'$ and the distance between the head and the ends in the Z-direction is $h$. The definitions of other quantities are unchanged here. For simplicity, we assume the rod to be smooth and the ratio $S'/L$ to be negligible. It means the helix part is much shorter than the total length of the rod, which corresponds to the initial state of the double helix formation process. Then the total potential energy can be modified as,
\begin{equation}
{E_{p}}=\frac{1}{2}B\ka^2 S'+\frac{1}{2}C\om^2L-T'h.
\end{equation}
Similarly, by using the variational method \eqref{energy minimization0}, we can obtain the equilibrium state (See the derivation in Appendix E). The results are shown in \reffig{fbp}. It can be found that the tension will make the pitch larger than that at equilibrium state. Furthermore, given the fact that real rods are extensible under tension, the pitch will become even larger than the theoretically predicated one. It is noteworthy that the $T'=0$ situation is exactly equivalent to the twist-slacking method. If we try to minimize $T'$ to eliminate the deviation of the pitch, $\be$ will increase and tend to $\pi/2$ as well. In this case, the twist-spinning method simply degenerates into the twist-slacking method. Therefore, our theory presented above for the twist-slacking method is still approximately valid even if a small $T'$ is performed to keep the double helix straight.

There is another type of widely used twist-spinning method. First the two strands are spun independently and then aligned to be parallel to each other. The adjacent ends of the two rods are linked. Then the torque and tension exerted on both ends are gradually removed to allow the double helix to form and reach an equilibrium state. Compared with the twist-slacking method, one of its disadvantages is that it can neither produce non-uniform double helix nor monitor and adjust the pitch variation for the newly formed double helix in real time.
\begin{figure}
  \centering
\begin{minipage}[t]{1\linewidth}
    \centering
    \includegraphics[width=0.5\textwidth]{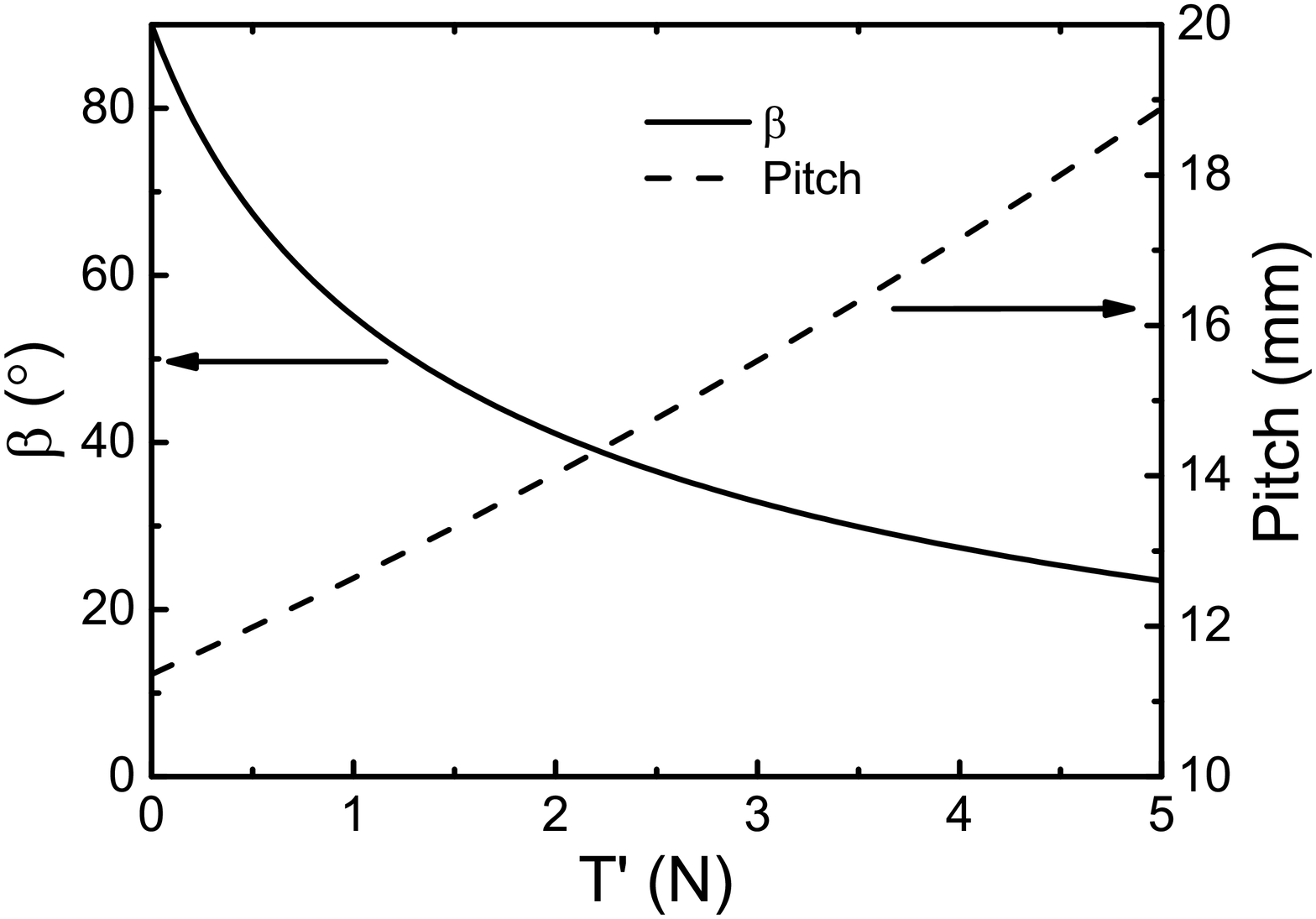}
\end{minipage}
  \caption{The $beta$ and pitch variations with the increase of the tension $T'$. Note the $T'=0$ situation exactly corresponds to the twist-slacking method.}\label{fbp}
\end{figure}

\section{\label{sec:level1}Conclusion}
In this article, we investigate the spontaneous formation of double helices for elastic rods under torsion. By using the variational method, the general equilibrium condition is obtained first. Then by applying further constraints, the stable configurations for both the smooth and the rough surface situations can be obtained. Specially, the self-contact and the surface friction are considered here. We find that non-uniform double helices can be formed due to the surface friction. Besides, the surface friction is determined by the specific process of how the rod is twisted and slacked. Based on this property, a method is proposed to produce double helices with designed configurations.
The experiments results indicate that our model can describe the formation of non-uniform double helices precisely. The method of producing designed double helices is proved to be effective and reliable and it can be used in manufacturing the electrical and mechanical devices based on the double helices of CNT yarns.

\begin{acknowledgements}
The authors would like to thank the Education Program for Talented Students of Xian Jiaotong University, National Natural Science Foundation of China (NSFC) (Grant No. 10704059), the Fundamental Research Funds for Central Universities (No. 2012jdgz04), the Scientific Research Foundation for the Returned Overseas Chinese Scholars, State Education Ministry (2013) and The College Students' Science and Technology Innovation Project of Xian Jiaotong University (2015).
\end{acknowledgements}

\renewcommand{\theequation}{A\arabic{equation}}
\setcounter{equation}{0}  
\section*{\label{sec:level1}Appendix A}
The exact curvature calculation for the curve defined by \equref{position vector} according to \equref{curvature} yields
\begin{equation}\label{curvature A}
\kappa_{ex}=\frac{4{\pi}^2 R \sqrt{{\left(P'\right)}^2 +{P}^2 +{\left(2\pi R\right)}^2}}{\left[ P^2 +{\left(2\pi R\right)}^2\right]}.
\end{equation}
where \(\kappa_{ex}\) denotes the exact curvature and \(P'\) denotes \(dP/d\theta\)
We define \(e=\kappa_{ex}/\kappa\), where \(\kappa\) is the approximated curvature calculated in \equref{Eq curvature1}. Then, we can obtain
\begin{equation}\label{error A}
e=\sqrt{1+\frac{1}{4\pi^2}\left(\frac{dP}{ds}\right)^2},
\end{equation}
where we use \equref{ds0} to rewrite \(P'\) as follows
\begin{equation}
\frac{dP}{d\theta}=\frac{dP}{ds} \sqrt{R^2+\left(\frac{P}{2\pi}\right)^2}.
\end{equation}

If we calculate $e$ for all the curves in \reffig{kzero:subfig}, we can find the typical range of $|e-1|$, namely the relative error, is
\begin{equation}\label{error}
|e-1|< 0.15\%.
\end{equation}
Thus the approximation used in \equref{Eq torsion1} is precise enough and valid.

\renewcommand{\theequation}{B\arabic{equation}}
\setcounter{equation}{0}  
\section*{\label{sec:level1}Appendix B}
Assume $\al(s)$ and $R(s)$ are added by infinitesimal functions $\delta \al(s)$ and $\delta R(s)$ respectively, the variation of the total potential energy can be expressed as
\begin{equation}\label{energy minimizationA}
\begin{aligned}
\delta E_p=& 2\int_0^{S/2}\left[\left(B\ka \frac{\partial \ka}{\partial \al}+C\om \frac{\partial \om}{\partial \al}+\frac{\pa w}{\pa \al}\right)\delta \al \right.\\
&\left.+\left(B\ka \frac{\partial \ka}{\partial R}+C\om \frac{\partial \om}{\partial R}+\frac{\pa w}{\pa R}\right)\delta R \right] ds\\
&+C\om_t(L-S)\delta\om_t \\
=&0.
\end{aligned}
\end{equation}
By utilizing the Calugareanu invariant, $\delta \om_t$ here can be expressed as
\begin{equation}\label{eq deltaom A}
\begin{aligned}
(L-S)\delta\om_t=&-2 \int_0^{S/2}\left[\left(\frac{\partial \ta}{\partial \al}+\frac{\partial \om}{\partial \al}\right)\delta \al \right.\\
&\left.+ \left(\frac{\partial \ta}{\partial R}+\frac{\partial \om}{\partial R}\right)\delta R \right] ds.\\
\end{aligned}
\end{equation}
Therefore, by plugging \equref{eq deltaom A} into \equref{energy minimizationA}, we can obtain
\begin{equation}\label{eq step A3}
\begin{aligned}
\delta E_p&= 2\int_0^{S/2} \left[ B\ka \frac{\partial \ka}{\partial \al}-C\om_t \frac{\partial \ta}{\partial \al}+\frac{\pa w}{\pa \al} \right.\\
&\left. +C(\om-\om_t)\frac{\partial \om}{\partial \al}\right]\delta \al \\
&+\left[ B\ka \frac{\partial \ka}{\partial R}-C\om_t \frac{\partial \ta}{\partial R}+\frac{\pa w}{\pa R} \right.\\
&\left. +C(\om-\om_t)\frac{\partial \om}{\partial \al}+\right]\delta \al ds \\
&=0.\\
\end{aligned}
\end{equation}
Since $\delta \al(s)$ and $\delta R(s)$ can be arbitrary infinitesimal functions, we can obtain
\begin{equation}\label{eq step A4}
\begin{split}
&B\ka \frac{\partial \ka}{\partial \al}-C\om_t \frac{\partial \ta}{\partial \al}+C(\om-\om_t)\frac{\partial \om}{\partial \al}+\frac{\pa w}{\pa \al} =0\\
&B\ka \frac{\partial \ka}{\partial R}-C\om_t \frac{\partial \ta}{\partial R}+C(\om-\om_t)\frac{\partial \om}{\partial R} +\frac{\pa w}{\pa R}=0\\
\end{split}
\end{equation}
Then let $s=S/2$, we can obtain
\begin{equation}\label{eq step A5}
\begin{split}
&\left.\left(B\ka \frac{\partial \ka}{\partial \al}-C\om \frac{\partial \ta}{\partial \al}+\frac{\pa w}{\pa \al}\right)\right|_{s=S/2}=0.\\
&\left.\left(B\ka \frac{\partial \ka}{\partial R}-C\om \frac{\partial \ta}{\partial R}+\frac{\pa w}{\pa R}\right)\right|_{s=S/2}=0,\\
\end{split}
\end{equation}
where we use the definition $\om_t=\om(S/2)$, as mentioned previously.
Since $S$ can take any value here, \equref{eq step A5} is equivalent to
\begin{equation}\label{eq dom/ds}
\om=\left.\left(B\ka\frac{\pa\ka}{\pa\al}+\frac{\pa w}{\pa\al}\right)\right/C\frac{\pa\ta}{\pa\al}
\end{equation}
and
\begin{equation}
C\frac{\pa\ta}{\pa R}\left(B\ka\frac{\pa\ka}{\pa\al}+\frac{\pa w}{\pa\al}\right)-C\frac{\pa\ta}{\pa\al}\left(B\ka\frac{\pa\ka}{\pa R}+\frac{\pa w}{\pa R}\right)=0
\end{equation}

\renewcommand{\theequation}{C\arabic{equation}}
\setcounter{equation}{0}  
\section*{\label{sec:level1}Appendix C}
We define
\begin{equation}
Tw=Tw_h+Tw_t,
\end{equation}
where
\begin{equation}\label{Eq Twh0}
Tw_h\equiv 2\cdot \frac{1}{2\pi} \int_0^{S/2} \omega_h ds,
\end{equation}
and
\begin{equation}\label{Eq Twt0}
Tw_t \equiv 2\cdot \frac{1}{2\pi} \int_{S/2}^{L/2} \omega_t ds=\frac{\left(L-S\right)\omega_t }{2\pi}.
\end{equation}
Then by using the Calugareanu invariant, we can obtain
\begin{equation}
\om_t=\frac{2\pi\left(N-Wr-Tw_h\right)}{L-S}.
\end{equation}
let $s=S/2$, \equref{eq dom/ds1} can be written as
\begin{equation}\label{Eq step1}
\om_t=\left.\frac{B}{C}\cdot\frac{\kappa d\kappa}{d\tau}\right|_{s=S/2}=\frac{2\pi\left(N-Wr-Tw_h\right)}{L-S}.
\end{equation}
Then differentiate the both sides of \equref{Eq step1} with respect to S and use the definition of \(Wr\) in \equref{Eq Lk=Tw+Wr} and \(Tw_h\) in \equref{Eq Twh0}, $Tw_h$ can be removed,
\begin{equation}\label{Eq step2}
\frac{B}{C}\cdot\frac{d}{dS}\left(\left.\frac{\kappa d\kappa}{d\tau}\right|_{s=S/2}\right)=\left.\frac{\ta -k(S)}{S-L}\right|_{s=S/2},
\end{equation}
where, as mentioned above,
\begin{equation}
k(S)=2\pi\frac{ dN(S)}{dS}.
\end{equation}
Since there is no explicit S-dependence in the expression of $\al$, \equref{Eq step2} can be equivalently expressed as
\begin{equation}\label{Eq step2eq}
\frac{B}{2C}\cdot\frac{d}{ds}\left(\frac{\kappa d\kappa}{d\tau}\right)=\frac{\ta -k(2s)}{2s-L}.
\end{equation}
By replacing $\kappa$ and $\tau$ with \equref{Eq curvature1} and \equref{Eq torsion1} respectively we can obtain
\begin{equation}\label{eq step4}
\frac{d\al}{ds}=\frac{C}{B(L-2s)}\cdot\frac{\left[\sin2\al-2Rk(2s)\right]\cos^22\al}{\cos^32\al -1}.
\end{equation}

Next,we need to determine the initial condition. For $S\rightarrow0$, \equref{Eq step1} turns into
\begin{equation}\label{Eq step3}
\left.\frac{B}{C}\cdot \frac{\kappa d\kappa}{d\tau}\right|_{s=0}=\frac{2\pi N}{L}.
\end{equation}
By replacing $\kappa$ and $\tau$ with \equref{Eq curvature1} and \equref{Eq torsion1} respectively we can obtain the initial condition for $\al$
\begin{equation}
\left.(1-\cos2\al)\tan2\al\right|_{s=0}=\frac{C}{B}\frac{4\pi RN}{L}.
\end{equation}

Then we can calculate the tension exerted by the rod. By using \equref{Eq Ep1}, \equref{force0} and \equref{eq dW}, we can obtain
\begin{equation}\label{eq step A6}
\begin{aligned}
T(S)&=-\frac{dE_p}{dS}+C\om_t k(S)\\
&=\left.\frac{1}{2}B\ka^2\right|_{s=S/2}+C\om_t\left[\frac{1}{L-S} \cdot\frac{d\om_t}{dS}+k(S)\right]\\
&=\left.\frac{1}{2}\left\lbrace B\ka^2+C\om \left[\frac{1}{L-S} \cdot\frac{d\om}{d\al}\cdot\frac{d\al}{ds}+2k(2s)\right]\right\rbrace\right|_{s=S/2},
\end{aligned}
\end{equation}
 in which we use the definition of $\om_t$ and simple substitution
\begin{equation}\label{eq step A7}
\frac{d\om_t}{dS}=\frac{d\om(S/2;S)}{dS}=\left.\frac{1}{2}\frac{d\om}{d\al}\cdot\frac{d\al}{ds}\right|_{s=S/2}.
\end{equation}

By using \equref{Eq curvature1}, \equref{Eq torsion1}, \equref{eq dom/ds1} and \equref{eq step4}, the tension can be expressed as
\begin{equation}
T(S)=\left.\frac{B}{8R^2}\cdot\frac{(1-\cos2\al)^2(2+\cos2\al)}{\cos2\al}\right|_{s=S/2}.
\end{equation}

\renewcommand{\theequation}{D\arabic{equation}}
\setcounter{equation}{0}  
\section*{\label{sec:level1}Appendix D}
Now, we assume the radius R is not a fixed parameter. For double helix in equilibrium, the configuration can be regarded as a function of R. Consider a piece of double helix with an infinitesimal length $\delta S$ in the whole double helix. Since $\delta S$ is infinitesimal, the double helix can be regarded as uniform here. We assume that this piece of double helix is isolated from the rest of the double helix. Therefore the Calugareanu invariant is valid for this piece of double helix and it can simplified as
\begin{equation}
Lk=\frac{1}{2\pi}(\om +\ta)\delta S;
\end{equation}
The total energy for this piece of rod can be expressed as
\begin{equation}
\delta E=(\frac{C\om^2}{2}+\frac{B\kappa^2}{2})\delta S.
\end{equation}
Then differentiate both sides with respect to R and we can obtain
\begin{equation}\label{eq ddeltaE}
d\delta E=(C\om\frac{d\om}{dR}+B\kappa\frac{d\ka}{dR})\delta SdR,
\end{equation}
where
\begin{equation}\label{eq dw/dr}
\begin{aligned}
\frac{d\om}{dR}&=-\frac{d\ta}{dR}=-\frac{\partial \ta}{\partial \al}\cdot \frac{d\al}{dR}-\frac{\partial \ta}{\partial R}\\
&=-\frac{\cos2\al}{R}\cdot\frac{d\al}{dR}+\frac{\sin2\al}{2R^2},
\end{aligned}
\end{equation}
and
\begin{equation}\label{eq dk/dr}
\begin{aligned}
\frac{d\ka}{dR}&=\frac{\partial \ka}{\partial \al}\cdot \frac{d\al}{dR}+\frac{\partial \ka}{\partial R}\\
&=\frac{\sin2\al}{R}\cdot \frac{d\al}{dR}-\frac{1-\cos2\al}{2R^2}.
\end{aligned}
\end{equation}
According the definition of $\al$ \equref{eq alpha def}, we have
\begin{equation}\label{eq da/dr}
\frac{d\al}{dR}=\frac{1-\cos2\al}{2}\cdot(\frac{\tan\al}{R}-\frac{\tan\al}{P}\cdot\frac{dP}{dR}).
\end{equation}
Plug \equref{Eq curvature1}, \equref{eq dom/ds1}, \equref{eq dw/dr}, \equref{eq dk/dr} and \equref{eq da/dr} into \equref{eq ddeltaE}, we can obtain
\begin{equation}
d\delta E=-\left[\frac{B}{4R^3}\cdot\frac{(1-\cos2\al)^2}{\cos2\al}\right]\delta SdR.
\end{equation}
Then the normal reaction between two strands for unit length can be expressed as
\begin{equation}
F=-\frac{1}{d\delta S}\cdot\frac{d\delta E}{dD}=\frac{B}{8R^3}\cdot\frac{(1-\cos2\al)^2}{\cos2\al},
\end{equation}
where $D=2R$ is the diameter of the double helix.

\renewcommand{\theequation}{E\arabic{equation}}
\setcounter{equation}{0}  
\section*{\label{sec:level1}Appendix E}
\begin{figure}
\begin{minipage}[t]{0.8\linewidth}
    \centering
    \includegraphics[width=1\textwidth]{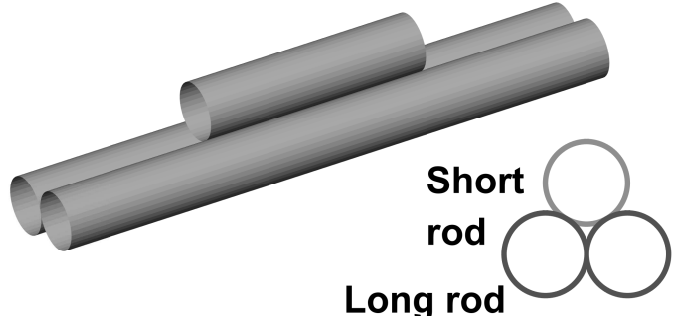}
\end{minipage}
\caption{Schematic of the method of measuring the coefficient of friction. The three rods attach each other. The cross section are shown in the right bottom.}\label{friction}
\end{figure}
A fraction of rubber rod ($L\approx2cm$) is put on two long parallel rubber rods with the same material and length. These two long rods are attached so that all the three rods attach each other, as illustrated in \reffig{friction}. Hold the ends of the long rods and then lift one end of them. Increase the angle of inclination gradually until the short rod start to slip. Then we can calculate $\mu$ from this critical angle. By using this method, the measured value from $\mu$ is between 0.4 and 0.6.

\renewcommand{\theequation}{F\arabic{equation}}
\setcounter{equation}{0}  
\section*{\label{sec:level1}Appendix F}
By utilizing the geometric relation, $h$ can be expressed in terms of $S'$ and $\be$ easily,
\begin{equation}\label{f1}
h=\frac{S'}{\cos\al}+\frac{L-S}{2\tan\be}.
\end{equation}
Furthermore, $\be$ can be expressed in terms of $S'$,
\begin{equation}\label{11}
\sin\be=\frac{L-S}{L-S'}.
\end{equation}
Note here we have two independent variables, namely, $S'$ and $\al$. 
By using \equref{energy minimization0}, we can obtain,
\begin{equation}\label{f2}
\frac{\partial E_p}{\partial S'}=0,
\end{equation}
and
\begin{equation}\label{f3}
\frac{\partial E_p}{\partial \al}=0.
\end{equation}
Since we have assumed that $S'/L$ is negligible, we can equate $S'/L$ with $0$ in \equref{f2} and \equref{f3}. Then the equations that determine $\al$ and $\be$ can be obtained,
\begin{equation}\label{f4}
B\ka\frac{\partial\ka}{\partial \al}-2\pi Cn \frac{\partial \ta}{\partial \al}+\frac{1}{2}F\sin\al=0,
\end{equation}
\begin{equation}\label{f5}
B\ka^2-4\pi Cn\ta-F(\cos\al-\sec\be)=0.
\end{equation}
where $n$ is defined by $N/L$.
Combined with \equref{Eq curvature1} and \equref{Eq torsion1}, \equref{f4} and \equref{f5} can be solved. The results are shown in \reffig{fbp}.

\bibliographystyle{apsrev4-1}
\bibliography{reference}

\begin{thebibliography}{39}%
\makeatletter
\providecommand \@ifxundefined [1]{%
 \@ifx{#1\undefined}
}%
\providecommand \@ifnum [1]{%
 \ifnum #1\expandafter \@firstoftwo
 \else \expandafter \@secondoftwo
 \fi
}%
\providecommand \@ifx [1]{%
 \ifx #1\expandafter \@firstoftwo
 \else \expandafter \@secondoftwo
 \fi
}%
\providecommand \natexlab [1]{#1}%
\providecommand \enquote  [1]{``#1''}%
\providecommand \bibnamefont  [1]{#1}%
\providecommand \bibfnamefont [1]{#1}%
\providecommand \citenamefont [1]{#1}%
\providecommand \href@noop [0]{\@secondoftwo}%
\providecommand \href [0]{\begingroup \@sanitize@url \@href}%
\providecommand \@href[1]{\@@startlink{#1}\@@href}%
\providecommand \@@href[1]{\endgroup#1\@@endlink}%
\providecommand \@sanitize@url [0]{\catcode `\\12\catcode `\$12\catcode
  `\&12\catcode `\#12\catcode `\^12\catcode `\_12\catcode `\%12\relax}%
\providecommand \@@startlink[1]{}%
\providecommand \@@endlink[0]{}%
\providecommand \url  [0]{\begingroup\@sanitize@url \@url }%
\providecommand \@url [1]{\endgroup\@href {#1}{\urlprefix }}%
\providecommand \urlprefix  [0]{URL }%
\providecommand \Eprint [0]{\href }%
\providecommand \doibase [0]{http://dx.doi.org/}%
\providecommand \selectlanguage [0]{\@gobble}%
\providecommand \bibinfo  [0]{\@secondoftwo}%
\providecommand \bibfield  [0]{\@secondoftwo}%
\providecommand \translation [1]{[#1]}%
\providecommand \BibitemOpen [0]{}%
\providecommand \bibitemStop [0]{}%
\providecommand \bibitemNoStop [0]{.\EOS\space}%
\providecommand \EOS [0]{\spacefactor3000\relax}%
\providecommand \BibitemShut  [1]{\csname bibitem#1\endcsname}%
\let\auto@bib@innerbib\@empty
\bibitem [{\citenamefont {Bednar}\ \emph {et~al.}(1994)\citenamefont {Bednar},
  \citenamefont {Furrer}, \citenamefont {Stasiak},\ and\ \citenamefont
  {Dubochet}}]{shape_of_supercoiled}%
  \BibitemOpen
  \bibfield  {author} {\bibinfo {author} {\bibfnamefont {J.}~\bibnamefont
  {Bednar}}, \bibinfo {author} {\bibfnamefont {P.}~\bibnamefont {Furrer}},
  \bibinfo {author} {\bibfnamefont {A.}~\bibnamefont {Stasiak}}, \ and\
  \bibinfo {author} {\bibfnamefont {J.}~\bibnamefont {Dubochet}},\ }\href@noop
  {} {\bibfield  {journal} {\bibinfo  {journal} {J. Mol. Biol}\ }\textbf
  {\bibinfo {volume} {235}},\ \bibinfo {pages} {825} (\bibinfo {year}
  {1994})}\BibitemShut {NoStop}%
\bibitem [{\citenamefont {Allemand}\ \emph {et~al.}(1998)\citenamefont
  {Allemand}, \citenamefont {Bensimon}, \citenamefont {Lavery},\ and\
  \citenamefont {Croquette}}]{twisting_DNA1}%
  \BibitemOpen
  \bibfield  {author} {\bibinfo {author} {\bibfnamefont {J.~F.}\ \bibnamefont
  {Allemand}}, \bibinfo {author} {\bibfnamefont {D.}~\bibnamefont {Bensimon}},
  \bibinfo {author} {\bibfnamefont {R.}~\bibnamefont {Lavery}}, \ and\ \bibinfo
  {author} {\bibfnamefont {V.}~\bibnamefont {Croquette}},\ }\href@noop {}
  {\bibfield  {journal} {\bibinfo  {journal} {Proc. Natl. Acad. Sci. U.S.A}\
  }\textbf {\bibinfo {volume} {95}},\ \bibinfo {pages} {14152} (\bibinfo {year}
  {1998})}\BibitemShut {NoStop}%
\bibitem [{\citenamefont {Strick}\ \emph
  {et~al.}(1999{\natexlab{a}})\citenamefont {Strick}, \citenamefont
  {Bensimon},\ and\ \citenamefont {Croquette}}]{twisting_DNA2}%
  \BibitemOpen
  \bibfield  {author} {\bibinfo {author} {\bibfnamefont {T.}~\bibnamefont
  {Strick}}, \bibinfo {author} {\bibfnamefont {D.}~\bibnamefont {Bensimon}}, \
  and\ \bibinfo {author} {\bibfnamefont {V.}~\bibnamefont {Croquette}},\
  }\href@noop {} {\bibfield  {journal} {\bibinfo  {journal} {Genetica}\
  }\textbf {\bibinfo {volume} {106}},\ \bibinfo {pages} {57} (\bibinfo {year}
  {1999}{\natexlab{a}})}\BibitemShut {NoStop}%
\bibitem [{\citenamefont {Strick}\ \emph {et~al.}(1998)\citenamefont {Strick},
  \citenamefont {Allemand}, \citenamefont {Bensimon},\ and\ \citenamefont
  {Croquette}}]{Strick1998}%
  \BibitemOpen
  \bibfield  {author} {\bibinfo {author} {\bibfnamefont {T.~R.}\ \bibnamefont
  {Strick}}, \bibinfo {author} {\bibfnamefont {J.~F.}\ \bibnamefont
  {Allemand}}, \bibinfo {author} {\bibfnamefont {D.}~\bibnamefont {Bensimon}},
  \ and\ \bibinfo {author} {\bibfnamefont {V.}~\bibnamefont {Croquette}},\
  }\href@noop {} {\bibfield  {journal} {\bibinfo  {journal} {Biophysical
  Journal}\ }\textbf {\bibinfo {volume} {74}},\ \bibinfo {pages} {2016}
  (\bibinfo {year} {1998})}\BibitemShut {NoStop}%
\bibitem [{\citenamefont {Strick}\ \emph {et~al.}(2003)\citenamefont {Strick},
  \citenamefont {Dessinges}, \citenamefont {Charvin}, \citenamefont {Dekker},
  \citenamefont {Allemand}, \citenamefont {Bensimon},\ and\ \citenamefont
  {Croquette}}]{Strick2003}%
  \BibitemOpen
  \bibfield  {author} {\bibinfo {author} {\bibfnamefont {T.~R.}\ \bibnamefont
  {Strick}}, \bibinfo {author} {\bibfnamefont {M.~N.}\ \bibnamefont
  {Dessinges}}, \bibinfo {author} {\bibfnamefont {G.}~\bibnamefont {Charvin}},
  \bibinfo {author} {\bibfnamefont {N.~H.}\ \bibnamefont {Dekker}}, \bibinfo
  {author} {\bibfnamefont {J.~F.}\ \bibnamefont {Allemand}}, \bibinfo {author}
  {\bibfnamefont {D.}~\bibnamefont {Bensimon}}, \ and\ \bibinfo {author}
  {\bibfnamefont {V.}~\bibnamefont {Croquette}},\ }\href@noop {} {\bibfield
  {journal} {\bibinfo  {journal} {Rep. Prog. Phys.}\ }\textbf {\bibinfo
  {volume} {66}},\ \bibinfo {pages} {1} (\bibinfo {year} {2003})}\BibitemShut
  {NoStop}%
\bibitem [{\citenamefont {Charvin}\ \emph {et~al.}(2005)\citenamefont
  {Charvin}, \citenamefont {Vologodskii}, \citenamefont {Bensimon},\ and\
  \citenamefont {Croquette}}]{braiding_DNA1}%
  \BibitemOpen
  \bibfield  {author} {\bibinfo {author} {\bibfnamefont {G.}~\bibnamefont
  {Charvin}}, \bibinfo {author} {\bibfnamefont {A.}~\bibnamefont
  {Vologodskii}}, \bibinfo {author} {\bibfnamefont {D.}~\bibnamefont
  {Bensimon}}, \ and\ \bibinfo {author} {\bibfnamefont {V.}~\bibnamefont
  {Croquette}},\ }\href@noop {} {\bibfield  {journal} {\bibinfo  {journal}
  {Biophysical Journal}\ }\textbf {\bibinfo {volume} {88}},\ \bibinfo {pages}
  {4124} (\bibinfo {year} {2005})}\BibitemShut {NoStop}%
\bibitem [{\citenamefont {Fojta}\ \emph {et~al.}(1998)\citenamefont {Fojta},
  \citenamefont {Stankova}, \citenamefont {Palecek}, \citenamefont
  {Koscielniak},\ and\ \citenamefont {Mitas}}]{biosensor}%
  \BibitemOpen
  \bibfield  {author} {\bibinfo {author} {\bibfnamefont {M.}~\bibnamefont
  {Fojta}}, \bibinfo {author} {\bibfnamefont {V.}~\bibnamefont {Stankova}},
  \bibinfo {author} {\bibfnamefont {E.}~\bibnamefont {Palecek}}, \bibinfo
  {author} {\bibfnamefont {P.}~\bibnamefont {Koscielniak}}, \ and\ \bibinfo
  {author} {\bibfnamefont {J.}~\bibnamefont {Mitas}},\ }\href@noop {}
  {\bibfield  {journal} {\bibinfo  {journal} {Talanta}\ }\textbf {\bibinfo
  {volume} {46}},\ \bibinfo {pages} {155} (\bibinfo {year} {1998})}\BibitemShut
  {NoStop}%
\bibitem [{\citenamefont {Fojta}\ \emph {et~al.}(1997)\citenamefont {Fojta},
  \citenamefont {Havran},\ and\ \citenamefont {Palecek}}]{biosensor2}%
  \BibitemOpen
  \bibfield  {author} {\bibinfo {author} {\bibfnamefont {M.}~\bibnamefont
  {Fojta}}, \bibinfo {author} {\bibfnamefont {L.}~\bibnamefont {Havran}}, \
  and\ \bibinfo {author} {\bibfnamefont {E.}~\bibnamefont {Palecek}},\
  }\href@noop {} {\bibfield  {journal} {\bibinfo  {journal} {Electroanalysis}\
  }\textbf {\bibinfo {volume} {9}},\ \bibinfo {pages} {1033} (\bibinfo {year}
  {1997})}\BibitemShut {NoStop}%
\bibitem [{\citenamefont {Shang}\ \emph {et~al.}(2013)\citenamefont {Shang},
  \citenamefont {Li}, \citenamefont {He}, \citenamefont {Du}, \citenamefont
  {Zhang}, \citenamefont {Shi}, \citenamefont {Wu}, \citenamefont {Li},
  \citenamefont {Li}, \citenamefont {Wei}, \citenamefont {Wang}, \citenamefont
  {Zhu}, \citenamefont {Wu},\ and\ \citenamefont {Cao}}]{CNT1}%
  \BibitemOpen
  \bibfield  {author} {\bibinfo {author} {\bibfnamefont {Y.}~\bibnamefont
  {Shang}}, \bibinfo {author} {\bibfnamefont {Y.}~\bibnamefont {Li}}, \bibinfo
  {author} {\bibfnamefont {X.}~\bibnamefont {He}}, \bibinfo {author}
  {\bibfnamefont {S.}~\bibnamefont {Du}}, \bibinfo {author} {\bibfnamefont
  {L.}~\bibnamefont {Zhang}}, \bibinfo {author} {\bibfnamefont
  {E.}~\bibnamefont {Shi}}, \bibinfo {author} {\bibfnamefont {S.}~\bibnamefont
  {Wu}}, \bibinfo {author} {\bibfnamefont {Z.}~\bibnamefont {Li}}, \bibinfo
  {author} {\bibfnamefont {P.}~\bibnamefont {Li}}, \bibinfo {author}
  {\bibfnamefont {J.}~\bibnamefont {Wei}}, \bibinfo {author} {\bibfnamefont
  {K.}~\bibnamefont {Wang}}, \bibinfo {author} {\bibfnamefont {H.}~\bibnamefont
  {Zhu}}, \bibinfo {author} {\bibfnamefont {D.}~\bibnamefont {Wu}}, \ and\
  \bibinfo {author} {\bibfnamefont {A.}~\bibnamefont {Cao}},\ }\href@noop {}
  {\bibfield  {journal} {\bibinfo  {journal} {ACS nano}\ }\textbf {\bibinfo
  {volume} {7}},\ \bibinfo {pages} {1446} (\bibinfo {year} {2013})}\BibitemShut
  {NoStop}%
\bibitem [{\citenamefont {Li}\ \emph {et~al.}(2013)\citenamefont {Li},
  \citenamefont {Shang}, \citenamefont {He}, \citenamefont {Peng},
  \citenamefont {Du}, \citenamefont {Shi}, \citenamefont {Wu}, \citenamefont
  {Li}, \citenamefont {Li},\ and\ \citenamefont {Cao}}]{CNT2}%
  \BibitemOpen
  \bibfield  {author} {\bibinfo {author} {\bibfnamefont {Y.}~\bibnamefont
  {Li}}, \bibinfo {author} {\bibfnamefont {Y.}~\bibnamefont {Shang}}, \bibinfo
  {author} {\bibfnamefont {X.}~\bibnamefont {He}}, \bibinfo {author}
  {\bibfnamefont {Q.}~\bibnamefont {Peng}}, \bibinfo {author} {\bibfnamefont
  {S.}~\bibnamefont {Du}}, \bibinfo {author} {\bibfnamefont {E.}~\bibnamefont
  {Shi}}, \bibinfo {author} {\bibfnamefont {S.}~\bibnamefont {Wu}}, \bibinfo
  {author} {\bibfnamefont {Z.}~\bibnamefont {Li}}, \bibinfo {author}
  {\bibfnamefont {P.}~\bibnamefont {Li}}, \ and\ \bibinfo {author}
  {\bibfnamefont {A.}~\bibnamefont {Cao}},\ }\href@noop {} {\bibfield
  {journal} {\bibinfo  {journal} {ACS nano}\ }\textbf {\bibinfo {volume} {7}},\
  \bibinfo {pages} {8128} (\bibinfo {year} {2013})}\BibitemShut {NoStop}%
\bibitem [{\citenamefont {Shang}\ \emph {et~al.}(2015)\citenamefont {Shang},
  \citenamefont {Wang}, \citenamefont {He}, \citenamefont {Li}, \citenamefont
  {Peng}, \citenamefont {Shi}, \citenamefont {Wang}, \citenamefont {Du},
  \citenamefont {Cao},\ and\ \citenamefont {Li}}]{CNT3}%
  \BibitemOpen
  \bibfield  {author} {\bibinfo {author} {\bibfnamefont {Y.}~\bibnamefont
  {Shang}}, \bibinfo {author} {\bibfnamefont {C.}~\bibnamefont {Wang}},
  \bibinfo {author} {\bibfnamefont {X.}~\bibnamefont {He}}, \bibinfo {author}
  {\bibfnamefont {J.}~\bibnamefont {Li}}, \bibinfo {author} {\bibfnamefont
  {Q.}~\bibnamefont {Peng}}, \bibinfo {author} {\bibfnamefont {E.}~\bibnamefont
  {Shi}}, \bibinfo {author} {\bibfnamefont {R.}~\bibnamefont {Wang}}, \bibinfo
  {author} {\bibfnamefont {S.}~\bibnamefont {Du}}, \bibinfo {author}
  {\bibfnamefont {A.}~\bibnamefont {Cao}}, \ and\ \bibinfo {author}
  {\bibfnamefont {Y.}~\bibnamefont {Li}},\ }\href@noop {} {\bibfield  {journal}
  {\bibinfo  {journal} {Nano energy}\ }\textbf {\bibinfo {volume} {12}},\
  \bibinfo {pages} {401} (\bibinfo {year} {2015})}\BibitemShut {NoStop}%
\bibitem [{\citenamefont {Gao}\ \emph {et~al.}(2016)\citenamefont {Gao},
  \citenamefont {Li}, \citenamefont {Li}, \citenamefont {Cheng}, \citenamefont
  {Wang}, \citenamefont {Wang},\ and\ \citenamefont {Li}}]{gao2016}%
  \BibitemOpen
  \bibfield  {author} {\bibinfo {author} {\bibfnamefont {L.}~\bibnamefont
  {Gao}}, \bibinfo {author} {\bibfnamefont {X.}~\bibnamefont {Li}}, \bibinfo
  {author} {\bibfnamefont {X.}~\bibnamefont {Li}}, \bibinfo {author}
  {\bibfnamefont {J.}~\bibnamefont {Cheng}}, \bibinfo {author} {\bibfnamefont
  {B.}~\bibnamefont {Wang}}, \bibinfo {author} {\bibfnamefont {Z.}~\bibnamefont
  {Wang}}, \ and\ \bibinfo {author} {\bibfnamefont {C.}~\bibnamefont {Li}},\
  }\href@noop {} {\bibfield  {journal} {\bibinfo  {journal} {RSC Adv}\ }\textbf
  {\bibinfo {volume} {6}},\ \bibinfo {pages} {57190} (\bibinfo {year}
  {2016})}\BibitemShut {NoStop}%
\bibitem [{\citenamefont {Haines}\ \emph {et~al.}(2014)\citenamefont {Haines},
  \citenamefont {Lima}, \citenamefont {N.~Li}, \citenamefont {Foroughi},
  \citenamefont {Madden}, \citenamefont {Kim}, \citenamefont {S.~Fang},
  \citenamefont {Goktepe}, \citenamefont {Goktepe}, \citenamefont {Mirvakili},
  \citenamefont {Naficy}, \citenamefont {Lepro}, \citenamefont {Oh},
  \citenamefont {Kozlov}, \citenamefont {Kim}, \citenamefont {Xu},
  \citenamefont {Swedlove}, \citenamefont {Wallace},\ and\ \citenamefont
  {Baughman}}]{Haines2014}%
  \BibitemOpen
  \bibfield  {author} {\bibinfo {author} {\bibfnamefont {C.~S.}\ \bibnamefont
  {Haines}}, \bibinfo {author} {\bibfnamefont {M.~D.}\ \bibnamefont {Lima}},
  \bibinfo {author} {\bibfnamefont {G.~M.~S.}\ \bibnamefont {N.~Li}}, \bibinfo
  {author} {\bibfnamefont {J.}~\bibnamefont {Foroughi}}, \bibinfo {author}
  {\bibfnamefont {J.~D.}\ \bibnamefont {Madden}}, \bibinfo {author}
  {\bibfnamefont {S.~H.}\ \bibnamefont {Kim}}, \bibinfo {author} {\bibfnamefont
  {M.~J. d.~A.}\ \bibnamefont {S.~Fang}}, \bibinfo {author} {\bibfnamefont
  {F.}~\bibnamefont {Goktepe}}, \bibinfo {author} {\bibfnamefont
  {O.}~\bibnamefont {Goktepe}}, \bibinfo {author} {\bibfnamefont {S.~M.}\
  \bibnamefont {Mirvakili}}, \bibinfo {author} {\bibfnamefont {S.}~\bibnamefont
  {Naficy}}, \bibinfo {author} {\bibfnamefont {X.}~\bibnamefont {Lepro}},
  \bibinfo {author} {\bibfnamefont {J.}~\bibnamefont {Oh}}, \bibinfo {author}
  {\bibfnamefont {M.~E.}\ \bibnamefont {Kozlov}}, \bibinfo {author}
  {\bibfnamefont {S.~J.}\ \bibnamefont {Kim}}, \bibinfo {author} {\bibfnamefont
  {X.}~\bibnamefont {Xu}}, \bibinfo {author} {\bibfnamefont {B.~J.}\
  \bibnamefont {Swedlove}}, \bibinfo {author} {\bibfnamefont {G.~G.}\
  \bibnamefont {Wallace}}, \ and\ \bibinfo {author} {\bibfnamefont {R.~H.}\
  \bibnamefont {Baughman}},\ }\href@noop {} {\bibfield  {journal} {\bibinfo
  {journal} {Science}\ }\textbf {\bibinfo {volume} {343}},\ \bibinfo {pages}
  {868} (\bibinfo {year} {2014})}\BibitemShut {NoStop}%
\bibitem [{\citenamefont {Yue}\ \emph {et~al.}(2015)\citenamefont {Yue},
  \citenamefont {Zhang}, \citenamefont {Yong}, \citenamefont {Zhou},\ and\
  \citenamefont {Zhou}}]{Yue2015}%
  \BibitemOpen
  \bibfield  {author} {\bibinfo {author} {\bibfnamefont {D.}~\bibnamefont
  {Yue}}, \bibinfo {author} {\bibfnamefont {X.}~\bibnamefont {Zhang}}, \bibinfo
  {author} {\bibfnamefont {H.}~\bibnamefont {Yong}}, \bibinfo {author}
  {\bibfnamefont {J.}~\bibnamefont {Zhou}}, \ and\ \bibinfo {author}
  {\bibfnamefont {Y.~H.}\ \bibnamefont {Zhou}},\ }\href@noop {} {\bibfield
  {journal} {\bibinfo  {journal} {Appl. Phys. Lett.}\ }\textbf {\bibinfo
  {volume} {107}},\ \bibinfo {pages} {111903} (\bibinfo {year}
  {2015})}\BibitemShut {NoStop}%
\bibitem [{\citenamefont {Goyal}\ \emph {et~al.}(2005)\citenamefont {Goyal},
  \citenamefont {Perkins},\ and\ \citenamefont
  {Lee}}]{loop_formation_DNA_Cable}%
  \BibitemOpen
  \bibfield  {author} {\bibinfo {author} {\bibfnamefont {S.}~\bibnamefont
  {Goyal}}, \bibinfo {author} {\bibfnamefont {N.~C.}\ \bibnamefont {Perkins}},
  \ and\ \bibinfo {author} {\bibfnamefont {C.~L.}\ \bibnamefont {Lee}},\
  }\href@noop {} {\bibfield  {journal} {\bibinfo  {journal} {Journal of
  Computational Physics}\ }\textbf {\bibinfo {volume} {209}},\ \bibinfo {pages}
  {371} (\bibinfo {year} {2005})}\BibitemShut {NoStop}%
\bibitem [{\citenamefont {Stump}\ \emph {et~al.}(1998)\citenamefont {Stump},
  \citenamefont {Fraser},\ and\ \citenamefont {Gates}}]{cable_DNA_writhe}%
  \BibitemOpen
  \bibfield  {author} {\bibinfo {author} {\bibfnamefont {D.~M.}\ \bibnamefont
  {Stump}}, \bibinfo {author} {\bibfnamefont {W.~B.}\ \bibnamefont {Fraser}}, \
  and\ \bibinfo {author} {\bibfnamefont {K.~E.}\ \bibnamefont {Gates}},\
  }\href@noop {} {\bibfield  {journal} {\bibinfo  {journal} {Proc. R. Soc.
  Lond. A}\ }\textbf {\bibinfo {volume} {454}},\ \bibinfo {pages} {2123}
  (\bibinfo {year} {1998})}\BibitemShut {NoStop}%
\bibitem [{\citenamefont {Goss}\ \emph {et~al.}(2005)\citenamefont {Goss},
  \citenamefont {Van Der~Heijden}, \citenamefont {Thompson},\ and\
  \citenamefont {Neukirch}}]{experiment1}%
  \BibitemOpen
  \bibfield  {author} {\bibinfo {author} {\bibfnamefont {V.~G.~A.}\
  \bibnamefont {Goss}}, \bibinfo {author} {\bibfnamefont {G.~H.~M.}\
  \bibnamefont {Van Der~Heijden}}, \bibinfo {author} {\bibfnamefont {J.~M.~T.}\
  \bibnamefont {Thompson}}, \ and\ \bibinfo {author} {\bibfnamefont
  {S.}~\bibnamefont {Neukirch}},\ }\href@noop {} {\bibfield  {journal}
  {\bibinfo  {journal} {Experimental Mechanics}\ }\textbf {\bibinfo {volume}
  {45}},\ \bibinfo {pages} {101} (\bibinfo {year} {2005})}\BibitemShut
  {NoStop}%
\bibitem [{\citenamefont {Champneys}\ and\ \citenamefont
  {Thompson}(1996)}]{experiment2}%
  \BibitemOpen
  \bibfield  {author} {\bibinfo {author} {\bibfnamefont {A.~R.}\ \bibnamefont
  {Champneys}}\ and\ \bibinfo {author} {\bibfnamefont {J.~M.~T.}\ \bibnamefont
  {Thompson}},\ }\href@noop {} {\bibfield  {journal} {\bibinfo  {journal}
  {Proc. R. Soc. Lond. A}\ }\textbf {\bibinfo {volume} {452}},\ \bibinfo
  {pages} {2467} (\bibinfo {year} {1996})}\BibitemShut {NoStop}%
\bibitem [{\citenamefont {Thompson}\ and\ \citenamefont
  {Champneys}(1996)}]{experiment3}%
  \BibitemOpen
  \bibfield  {author} {\bibinfo {author} {\bibfnamefont {J.~M.~T.}\
  \bibnamefont {Thompson}}\ and\ \bibinfo {author} {\bibfnamefont {A.~R.}\
  \bibnamefont {Champneys}},\ }\href@noop {} {\bibfield  {journal} {\bibinfo
  {journal} {Proc. R. Soc. Lond. A}\ }\textbf {\bibinfo {volume} {452}},\
  \bibinfo {pages} {117} (\bibinfo {year} {1996})}\BibitemShut {NoStop}%
\bibitem [{\citenamefont {Goriely}\ and\ \citenamefont
  {Tabor}(1998)}]{theory_Kirchhoff}%
  \BibitemOpen
  \bibfield  {author} {\bibinfo {author} {\bibfnamefont {A.}~\bibnamefont
  {Goriely}}\ and\ \bibinfo {author} {\bibfnamefont {M.}~\bibnamefont
  {Tabor}},\ }\href@noop {} {\bibfield  {journal} {\bibinfo  {journal} {Proc.
  R. Soc. Lond. A}\ }\textbf {\bibinfo {volume} {454}},\ \bibinfo {pages}
  {3183} (\bibinfo {year} {1998})}\BibitemShut {NoStop}%
\bibitem [{\citenamefont {Purohit}(2008)}]{theorypurohit}%
  \BibitemOpen
  \bibfield  {author} {\bibinfo {author} {\bibfnamefont {P.~K.}\ \bibnamefont
  {Purohit}},\ }\href@noop {} {\bibfield  {journal} {\bibinfo  {journal}
  {Journal of the Mechanics and Physics of Solids}\ }\textbf {\bibinfo {volume}
  {56}},\ \bibinfo {pages} {1715} (\bibinfo {year} {2008})}\BibitemShut
  {NoStop}%
\bibitem [{\citenamefont {Thompson}\ \emph {et~al.}(2002)\citenamefont
  {Thompson}, \citenamefont {van~der Heijden},\ and\ \citenamefont
  {Neukirch}}]{self_contact1}%
  \BibitemOpen
  \bibfield  {author} {\bibinfo {author} {\bibfnamefont {J.~M.~T.}\
  \bibnamefont {Thompson}}, \bibinfo {author} {\bibfnamefont {G.~H.~M.}\
  \bibnamefont {van~der Heijden}}, \ and\ \bibinfo {author} {\bibfnamefont
  {S.}~\bibnamefont {Neukirch}},\ }\href@noop {} {\bibfield  {journal}
  {\bibinfo  {journal} {Proc. R. Soc. Lond. A}\ }\textbf {\bibinfo {volume}
  {458}},\ \bibinfo {pages} {959} (\bibinfo {year} {2002})}\BibitemShut
  {NoStop}%
\bibitem [{\citenamefont {Tobias}\ \emph {et~al.}(2000)\citenamefont {Tobias},
  \citenamefont {Swigon},\ and\ \citenamefont {Coleman}}]{self_contact2}%
  \BibitemOpen
  \bibfield  {author} {\bibinfo {author} {\bibfnamefont {I.}~\bibnamefont
  {Tobias}}, \bibinfo {author} {\bibfnamefont {D.}~\bibnamefont {Swigon}}, \
  and\ \bibinfo {author} {\bibfnamefont {B.~D.}\ \bibnamefont {Coleman}},\
  }\href@noop {} {\bibfield  {journal} {\bibinfo  {journal} {Phys. Rev. E}\
  }\textbf {\bibinfo {volume} {61}},\ \bibinfo {pages} {747} (\bibinfo {year}
  {2000})}\BibitemShut {NoStop}%
\bibitem [{\citenamefont {Coleman}\ \emph {et~al.}(2000)\citenamefont
  {Coleman}, \citenamefont {Swigon},\ and\ \citenamefont
  {Tobias}}]{self_contact3}%
  \BibitemOpen
  \bibfield  {author} {\bibinfo {author} {\bibfnamefont {B.~D.}\ \bibnamefont
  {Coleman}}, \bibinfo {author} {\bibfnamefont {D.}~\bibnamefont {Swigon}}, \
  and\ \bibinfo {author} {\bibfnamefont {I.}~\bibnamefont {Tobias}},\
  }\href@noop {} {\bibfield  {journal} {\bibinfo  {journal} {Phys. Rev. E}\
  }\textbf {\bibinfo {volume} {61}},\ \bibinfo {pages} {759} (\bibinfo {year}
  {2000})}\BibitemShut {NoStop}%
\bibitem [{\citenamefont {Coleman}\ and\ \citenamefont
  {Swigon}(2000)}]{self_contact4}%
  \BibitemOpen
  \bibfield  {author} {\bibinfo {author} {\bibfnamefont {B.~D.}\ \bibnamefont
  {Coleman}}\ and\ \bibinfo {author} {\bibfnamefont {D.}~\bibnamefont
  {Swigon}},\ }\href@noop {} {\bibfield  {journal} {\bibinfo  {journal}
  {Journal of Elasticity}\ }\textbf {\bibinfo {volume} {60}},\ \bibinfo {pages}
  {173} (\bibinfo {year} {2000})}\BibitemShut {NoStop}%
\bibitem [{\citenamefont {Stricka}\ \emph {et~al.}(2000)\citenamefont
  {Stricka}, \citenamefont {Allemanda}, \citenamefont {Croquettea},\ and\
  \citenamefont {Bensimona}}]{chainmodel}%
  \BibitemOpen
  \bibfield  {author} {\bibinfo {author} {\bibfnamefont {T.}~\bibnamefont
  {Stricka}}, \bibinfo {author} {\bibfnamefont {J.~F.}\ \bibnamefont
  {Allemanda}}, \bibinfo {author} {\bibfnamefont {V.}~\bibnamefont
  {Croquettea}}, \ and\ \bibinfo {author} {\bibfnamefont {D.}~\bibnamefont
  {Bensimona}},\ }\href@noop {} {\bibfield  {journal} {\bibinfo  {journal}
  {Progress in Biophysics and Molecular Biology}\ }\textbf {\bibinfo {volume}
  {74}},\ \bibinfo {pages} {115} (\bibinfo {year} {2000})}\BibitemShut
  {NoStop}%
\bibitem [{\citenamefont {Strick}\ \emph
  {et~al.}(1999{\natexlab{b}})\citenamefont {Strick}, \citenamefont
  {Bensimon},\ and\ \citenamefont {Croquette}}]{DNAtorsion}%
  \BibitemOpen
  \bibfield  {author} {\bibinfo {author} {\bibfnamefont {T.}~\bibnamefont
  {Strick}}, \bibinfo {author} {\bibfnamefont {D.}~\bibnamefont {Bensimon}}, \
  and\ \bibinfo {author} {\bibfnamefont {V.}~\bibnamefont {Croquette}},\
  }\href@noop {} {\bibfield  {journal} {\bibinfo  {journal} {Genetica}\
  }\textbf {\bibinfo {volume} {106}},\ \bibinfo {pages} {57} (\bibinfo {year}
  {1999}{\natexlab{b}})}\BibitemShut {NoStop}%
\bibitem [{\citenamefont {Bouchiat}\ and\ \citenamefont
  {M\'{e}zard}(1998)}]{elasticmodel}%
  \BibitemOpen
  \bibfield  {author} {\bibinfo {author} {\bibfnamefont {C.}~\bibnamefont
  {Bouchiat}}\ and\ \bibinfo {author} {\bibfnamefont {M.}~\bibnamefont
  {M\'{e}zard}},\ }\href@noop {} {\bibfield  {journal} {\bibinfo  {journal}
  {Phys. Rev. Lett.}\ }\textbf {\bibinfo {volume} {80}},\ \bibinfo {pages}
  {1556} (\bibinfo {year} {1998})}\BibitemShut {NoStop}%
\bibitem [{\citenamefont {Przyby}\ and\ \citenamefont
  {Pieranski}(2001)}]{rope_geometry}%
  \BibitemOpen
  \bibfield  {author} {\bibinfo {author} {\bibfnamefont {S.}~\bibnamefont
  {Przyby}}\ and\ \bibinfo {author} {\bibfnamefont {P.}~\bibnamefont
  {Pieranski}},\ }\href@noop {} {\bibfield  {journal} {\bibinfo  {journal}
  {Eur. Phys. J. E}\ }\textbf {\bibinfo {volume} {4}},\ \bibinfo {pages} {445}
  (\bibinfo {year} {2001})}\BibitemShut {NoStop}%
\bibitem [{\citenamefont {Moffatt}\ and\ \citenamefont
  {Ricca}(1992)}]{calugareanu_invariant}%
  \BibitemOpen
  \bibfield  {author} {\bibinfo {author} {\bibfnamefont {H.~K.}\ \bibnamefont
  {Moffatt}}\ and\ \bibinfo {author} {\bibfnamefont {R.~L.}\ \bibnamefont
  {Ricca}},\ }\href@noop {} {\bibfield  {journal} {\bibinfo  {journal} {Proc.
  R. Soc. Lond. A}\ }\textbf {\bibinfo {volume} {439}},\ \bibinfo {pages} {411}
  (\bibinfo {year} {1992})}\BibitemShut {NoStop}%
\bibitem [{\citenamefont {White}(1969)}]{Self_link}%
  \BibitemOpen
  \bibfield  {author} {\bibinfo {author} {\bibfnamefont {J.}~\bibnamefont
  {White}},\ }\href@noop {} {\bibfield  {journal} {\bibinfo  {journal} {Am. J.
  Math.}\ }\textbf {\bibinfo {volume} {91}},\ \bibinfo {pages} {693} (\bibinfo
  {year} {1969})}\BibitemShut {NoStop}%
\bibitem [{\citenamefont {Fuller}(1971)}]{Writhing_number}%
  \BibitemOpen
  \bibfield  {author} {\bibinfo {author} {\bibfnamefont {F.~B.}\ \bibnamefont
  {Fuller}},\ }\href@noop {} {\bibfield  {journal} {\bibinfo  {journal} {Proc.
  Nant. Acad. Sci. U.S.A}\ }\textbf {\bibinfo {volume} {68}},\ \bibinfo {pages}
  {815} (\bibinfo {year} {1971})}\BibitemShut {NoStop}%
\bibitem [{\citenamefont {Michael}\ and\ \citenamefont
  {Thompson}(2008)}]{Simplification_Tw}%
  \BibitemOpen
  \bibfield  {author} {\bibinfo {author} {\bibfnamefont {J.}~\bibnamefont
  {Michael}}\ and\ \bibinfo {author} {\bibfnamefont {T.}~\bibnamefont
  {Thompson}},\ }\href@noop {} {\bibfield  {journal} {\bibinfo  {journal}
  {Proc. R. Soc. A}\ }\textbf {\bibinfo {volume} {464}},\ \bibinfo {pages}
  {2811} (\bibinfo {year} {2008})}\BibitemShut {NoStop}%
\bibitem [{\citenamefont {Klapper}(1996)}]{energy}%
  \BibitemOpen
  \bibfield  {author} {\bibinfo {author} {\bibfnamefont {I.}~\bibnamefont
  {Klapper}},\ }\href@noop {} {\bibfield  {journal} {\bibinfo  {journal}
  {Journal of Computationla Physics}\ }\textbf {\bibinfo {volume} {125}},\
  \bibinfo {pages} {325} (\bibinfo {year} {1996})}\BibitemShut {NoStop}%
\bibitem [{\citenamefont {Salerno}\ \emph {et~al.}(2012)\citenamefont
  {Salerno}, \citenamefont {Tempestini}, \citenamefont {Mai}, \citenamefont
  {Brogioli}, \citenamefont {Ziano}, \citenamefont {Cassina},\ and\
  \citenamefont {Mantegazza}}]{Salerno2012}%
  \BibitemOpen
  \bibfield  {author} {\bibinfo {author} {\bibfnamefont {D.}~\bibnamefont
  {Salerno}}, \bibinfo {author} {\bibfnamefont {A.}~\bibnamefont {Tempestini}},
  \bibinfo {author} {\bibfnamefont {I.}~\bibnamefont {Mai}}, \bibinfo {author}
  {\bibfnamefont {D.}~\bibnamefont {Brogioli}}, \bibinfo {author}
  {\bibfnamefont {R.}~\bibnamefont {Ziano}}, \bibinfo {author} {\bibfnamefont
  {V.}~\bibnamefont {Cassina}}, \ and\ \bibinfo {author} {\bibfnamefont
  {F.}~\bibnamefont {Mantegazza}},\ }\href@noop {} {\bibfield  {journal}
  {\bibinfo  {journal} {Phys. Rev. Lett.}\ }\textbf {\bibinfo {volume} {109}},\
  \bibinfo {pages} {118303} (\bibinfo {year} {2012})}\BibitemShut {NoStop}%
\bibitem [{\citenamefont {Marco}\ and\ \citenamefont
  {Siggia}(1994)}]{Marco1994}%
  \BibitemOpen
  \bibfield  {author} {\bibinfo {author} {\bibfnamefont {J.}~\bibnamefont
  {Marco}}\ and\ \bibinfo {author} {\bibfnamefont {E.}~\bibnamefont {Siggia}},\
  }\href@noop {} {\bibfield  {journal} {\bibinfo  {journal} {Science}\ }\textbf
  {\bibinfo {volume} {265}},\ \bibinfo {pages} {506} (\bibinfo {year}
  {1994})}\BibitemShut {NoStop}%
\bibitem [{\citenamefont {Moroz}\ and\ \citenamefont
  {Nelson}(1997)}]{Moroz1997}%
  \BibitemOpen
  \bibfield  {author} {\bibinfo {author} {\bibfnamefont {J.}~\bibnamefont
  {Moroz}}\ and\ \bibinfo {author} {\bibfnamefont {P.}~\bibnamefont {Nelson}},\
  }\href@noop {} {\bibfield  {journal} {\bibinfo  {journal} {Proc. Natl. Acad.
  Sci. USA}\ }\textbf {\bibinfo {volume} {94}},\ \bibinfo {pages} {14418}
  (\bibinfo {year} {1997})}\BibitemShut {NoStop}%
\bibitem [{\citenamefont {Vologodskii}\ \emph {et~al.}(1992)\citenamefont
  {Vologodskii}, \citenamefont {Levene}, \citenamefont {Klenin}, \citenamefont
  {Frank-Kamenetskii},\ and\ \citenamefont {Cozzarelli}}]{Vologodskii1992}%
  \BibitemOpen
  \bibfield  {author} {\bibinfo {author} {\bibfnamefont {A.~V.}\ \bibnamefont
  {Vologodskii}}, \bibinfo {author} {\bibfnamefont {S.~D.}\ \bibnamefont
  {Levene}}, \bibinfo {author} {\bibfnamefont {K.~V.}\ \bibnamefont {Klenin}},
  \bibinfo {author} {\bibfnamefont {M.}~\bibnamefont {Frank-Kamenetskii}}, \
  and\ \bibinfo {author} {\bibfnamefont {N.~R.}\ \bibnamefont {Cozzarelli}},\
  }\href@noop {} {\bibfield  {journal} {\bibinfo  {journal} {J. Mol. Biol.}\
  }\textbf {\bibinfo {volume} {227}},\ \bibinfo {pages} {1224} (\bibinfo {year}
  {1992})}\BibitemShut {NoStop}%
\bibitem [{\citenamefont {Wolf}\ and\ \citenamefont
  {Descamps}(1999)}]{possion2}%
  \BibitemOpen
  \bibfield  {author} {\bibinfo {author} {\bibfnamefont {A.~T.}\ \bibnamefont
  {Wolf}}\ and\ \bibinfo {author} {\bibfnamefont {P.}~\bibnamefont
  {Descamps}},\ }in\ \href@noop {} {\emph {\bibinfo {booktitle} {Performance of
  Exterior Building Wall}}},\ \bibinfo {editor} {edited by\ \bibinfo {editor}
  {\bibfnamefont {P.~G.}\ \bibnamefont {Johnson}}}\ (\bibinfo  {publisher}
  {ASTM International},\ \bibinfo {address} {West Conshohocken, PA},\ \bibinfo
  {year} {1999})\BibitemShut {NoStop}%
\end{thebibliography}%
\end{document}